\documentclass[twocolumn,trackchanges]{aastex631}
\usepackage{amsmath}

\newcommand{\kms}{${\rm km\,s}^{-1}$}
\newcommand{\mpc}{$h^{-1}{\rm Mpc}\,$}

\newcommand{\kpc}{$h^{-1}{\rm kpc}\,$}

\newcommand{\mfp}{$\lambda^{912}_{\rm mfp}$}
\newcommand{\MFP}{\lambda^{912}_{\rm mfp}}

\newcommand{\Myr}{\text {Myr }}

\newcommand{\zre}{$z_{\rm{re}}\,$}
\newcommand{\comment}[1]{}

\newcommand{\NHI}{\ensuremath{N_{\rm HI}\,}}

\newcommand{\taui}{\ensuremath{\tau_{\rm 912}\,}}

\newcommand{\GammaHI}{$\Gamma_{\rm -12}\,$}

\newcommand{\HI}{\hbox{H$\,\rm \scriptstyle I\ $}}
\def\H{\hbox{H}}

\hypersetup{draft}

\shorttitle{Hydrodynamic Response to Reionization II}
\shortauthors{Nasir et al.}

\graphicspath{{./}{}}

\begin{document}

\title{Hydrodynamic Response of the Intergalactic Medium to Reionization II: Physical Characteristics and Dynamics of Ionizing Photon Sinks}

\author{Fahad Nasir}
\affiliation{Department of Physics and Astronomy, University of California, Riverside, CA 92521, USA}

\author{Christopher Cain}
\affiliation{Department of Physics and Astronomy, University of California, Riverside, CA 92521, USA}

\author{Anson D'Aloisio}
\affiliation{Department of Physics and Astronomy, University of California, Riverside, CA 92521, USA}

\author{Nakul Gangolli}
\affiliation{Department of Physics and Astronomy, University of California, Riverside, CA 92521, USA}

\author{Matthew McQuinn}
\affiliation{Department of Astronomy, University of Washington, Seattle, WA 98195-1580, USA}

\begin{abstract}

\citet{2021arXiv210316610B} measured the mean free path of Lyman limit photons in the IGM at $z=6$.  The short value suggests that absorptions may have played a prominent role in reionization.   Here we study physical properties of ionizing photon sinks in the wake of ionization fronts (I-fronts) using radiative hydrodynamic simulations.   We quantify the contributions of gaseous structures to the Lyman limit opacity by tracking the column density distributions in our simulations.  Within $\Delta t = 10$ Myr of I-front passage, we find that self-shielding systems ($N_{\rm HI} > 10^{17.2}$ cm$^{-2}$) are comprised of two distinct populations: (1) over-density $\Delta \sim 50$ structures in photo-ionization equilibrium with the ionizing background; (2) $\Delta \gtrsim 100$ density peaks with fully neutral cores.  The self-shielding systems contribute more than half of the opacity at these times, but the IGM evolves considerably in $\Delta t \sim 100$ Myr as structures are flattened by pressure smoothing and photoevaporation.   By $\Delta t = 300$ Myr, they contribute $\lesssim 10 \%$ to the opacity in an average 1 Mpc$^3$ patch of the Universe. The percentage can be a factor of a few larger in over-dense patches, where more self-shielding systems survive.  We quantify the characteristic masses and sizes of self-shielding structures. Shortly after I-front passage, we find $M=10^{4} - 10^8$ M$_\odot$ and effective diameters $d_{\rm eff} = 1 - 20$ ckpc$/h$.  These scales increase as the gas relaxes. The picture herein presented may be different in dark matter models with suppressed small-scale power.      
\end{abstract}

\keywords{Intergalactic medium(813) --- Reionization(1383) --- Radiative transfer simulations(1967)}

\section{Introduction} \label{sec:intro}

Hubble Space Telescope (HST) measurements of the rest-frame ultraviolet (UV) luminosity function of $z>6$ galaxies have provided a first census of reionization sources \citep[e.g.][]{2015ApJ...810...71F, Bowens2015}.  These observations give us a broad view of plausible reionization histories that are consistent with the observed star formation history of the Universe \citep[e.g.][]{Robertson_2015ApJ, 2019ApJ...879...36F, 2021arXiv210207775B}.  Of equal importance to our understanding of reionization, however, are the sinks of ionizing photons. The sinks shaped how ionization fronts (I-fronts) progressed \citep{2005ApJ...624..491I} and they played an important role in setting the ionizing photon budget required to complete and maintain reionization \citep{Park2016,Daloisio2020}.

The sinks can be characterized by their distribution of \HI\ column densities. At $z=2-5$ (post-reionization), the column density distribution has been constrained by numerous quasar absorption spectrum studies \citep{ Storrie-Lombardi1994, Songaila2010, Prochaska2010,Rudie2013, Kim2013, Crighton2019}.  At these redshifts, much of the opacity is contributed by optically thick absorbers with columns $10^{17.2} < N_{\rm HI} < 10^{19}$ cm$^{-2}$, the so-called Lyman-limit systems (LLSs). For example, \citet{Prochaska2010} found that $\approx 55 \%$ of the Lyman limit opacity at $z=3.7$ is produced by $N_{\rm HI} \geq 10^{17.5}$ cm$^{-2}$ LLSs.   Although the exact nature of LLSs is debated, cosmological radiative transfer (RT) simulations successfully reproduce their observed abundance, and suggest that the LLSs correspond to $\Delta_g \sim 100$ gas at the outskirts of halos, where $\Delta_g$ is the gas density in units of the cosmic mean \citep{McQuinn2011b,Altay2011,Altay2013}. 

In contrast to this relatively well-studied picture of the post-reionization IGM, the properties of the sinks during reionization are poorly understood. This owes to a lack of observational constraints at $z>6$ as well as the computational challenges in simulating the sinks. On the theoretical side, the simulations of \citet{Park2016} and \citet[][Paper I]{Daloisio2020} suggest that the LyC opacity during reionization was considerably more complicated than at lower redshifts. Before a patch of the IGM was reionized, the gas clumped on a hierarchy of scales down to its Jeans mass, which could have been as low as $M_J \sim 10^4$ M$_\odot$ for gas at temperature $T\sim 10$ K. After an I-front swept through a region, the photoionized gas expanded in response to the sudden pressure increase, a process that we term relaxation. In addition, I-fronts became stuck within gaseous halos until they were photoevaporated \citep{Shapiro2004, Iliev2005}.  These processes were limited by the sound speed in the ionized gas ($c_s \sim 20$ km s$^{-1}$), so they took place over several hundred Myr.  During this period the LyC opacity evolved significantly.  Moreover, the non-trivial interplay between the hydrodynamic response and self-shielding renders the evolution dependent on the local intensity of the ionizing background, gas density, and redshift of reionization.

On the observational side, \citet{2021arXiv210316610B} recently extended to $z=6$ direct measurements of the mean free path (MFP) from stacked quasar absorption spectra.  At $z=5.1$ they found $\MFP = 37.71^{+5.31}_{-6.64}$ cMpc$/h$, consistent with the previous measurement of \citet{Worseck2014}.  At $z=6$, the short value of $\MFP = 3.57^{+3.09}_{-2.14}$ cMpc$/h$ measured by \citet{2021arXiv210316610B} implies a rapid evolution of the IGM opacity between $z=5-6$. \citet{2021arXiv210510511C} used radiative transfer (RT) simulations, applied with a new sub-grid model for the sinks, to argue that the rapid evolution favors a late and rapid reionization process driven by faint galaxies.   \citet{2021arXiv210510518D} assessed that a cumulative source output of $6.1^{+11}_{-2.4}$ ionizing photons per baryon are required for consistency with both the short $\MFP(z=6)$ and upper limits on the IGM neutral fraction from \citet{McGreer2015}.  Both papers concluded that the sinks played a principal role in shaping reionization if $\MFP(z=6)$ is as low as the \citet{2021arXiv210316610B} measurement.\footnote{At face value, the ionizing photon budgets in the models of \citet{2021arXiv210510511C}, $\approx3$ photons per H atom to complete reionization, appear discrepant with the larger budget of $6.1^{+11}_{-2.4}$ found by \citet{2021arXiv210510518D}.  However, the models of \citet{2021arXiv210510511C} allow $\MFP(z=6)$ to be 1$\sigma$ larger than the central value measured by \citet{2021arXiv210316610B}.  And they have global neutral fractions $\approx 20\%$ at $z=6$, in $\sim 2\sigma$ tension with the dark pixel constraints of \citet{McGreer2015}. If the calculations of \citet{2021arXiv210510518D} are adjusted for these allowances, the agreement is quite good.  Under these assumptions, they find a cumulative budget of $2.3$ photons per baryon at $z=6$, compared to 2.2 in the model of \citet{2021arXiv210510511C}. }    

In this paper, we use the suite of radiative hydrodynamics simulations published in Paper I to take a more in-depth look at the absorption systems responsible for the LyC opacity during reionization.   We will address four main questions: (1) What are the physical properties of the absorption systems that set the opacity?; (2) How do these properties depend on the local environment?;  (3) What are the length and mass scales that characterize these systems?; (4) How do the systems evolve during the relaxation process?  A detailed understanding of the sinks' physical nature will provide insight into what absorption systems set the ionizing photon budget for reionization.  It will also help us understand the early evolution of the LyC opacity and how it fits with the standard model of the post-reionization IGM. Lastly, the sinks are expected to play an important role in setting the spatial structure of reionization \citep{Miralda-Escude_2000ApJ,Furlanetto_2005MNRAS,McQuinn2007,Mao2019}.  Since almost all reionization observables are sensitive at some level to its morphology, an accurate model for the sinks is likely critical for confronting simulations with forthcoming observations.  One aim of this work is to inform the further development of such models \citep[e.g.][]{2021arXiv210510511C}.  

The structure of this paper is as follows. In \S2 we briefly summarize the simulations of Paper I. In \S \ref{sec:photoion}, we discuss the relationship between photo-ionization rate and neutral hydrogen density. Section \ref{sec:LyC_opacity} lays out contributions to the LyC opacity of the IGM.  In \S \ref{sec:demographics}, we explore demographics of optically thick absorbers.  Section \ref{sec:3d_sys} quantities physical properties of the structures responsible for optically thick absorbers.  We conclude in \S \ref{sec:conclusions}. Throughout we adopt a $\Lambda$CDM cosmology with $\Omega_{\rm m} = 0.31$,  $\Omega_{\rm b} = 0.048$, $H_{\rm 0}= 100h\,$\kms$\,{\rm Mpc}^{-1}$, with $h = 0.68$, $\sigma_{\rm 8} = 0.82$, $n_{\rm s} = 0.9667$ and a hydrogen mass fraction of $X_{\rm Hy} =0.7547$, consistent with the latest measurements \citep{Planck2018}. Proper distances are denoted with a ``p" prefix (e.g. pMpc), while all other distances are reported in comoving units.

\section{Radiative Hydrodynamics simulations}\label{sec:sim}

We briefly summarize the main features of the fully coupled RT hydrodynamic simulations from Paper I. The simulations were run using a modified version of the RadHydro code, which combines ray tracing RT with Eulerian hydrodynamics \citep{Trac_2004NewA, Trac2007, Trac_2008ApJ}. All of the runs have $N={1024}^3$ dark matter particles, gas cells and RT cells, with box sizes of $L=1.024$~\mpc. This provides a gas/RT cell width of 1 \kpc, small enough to capture reasonably the pre-reionization Jeans scale of the gas. In Appendix \ref{app:convergence}, we present a series of numerical convergence tests (see also Appendix A of Paper I).  We will discuss these tests in the relevant contexts below. 

The simulations do not explicitly model the sources of reionization.  Instead, the gas is ionized by external sources, a process that we model by sending plane parallel I-fronts through the simulation volume.   The box is divided into $N_{\rm dom} = {32}^3$ cubic RT domains. Source cells are placed on two adjacent sides of each domain and they are turned on at a specified redshift of reionization, $z_{\rm re}$ (i.e. rays are sent from two directions within each domain).  As discussed in Paper I, this setup affords us a clean interpretation of the dynamics; the gas is ionized at nearly the same time and at the same impinging radiation intensity.   Appendix A of Paper I demonstrates that the domain setup does not significantly alter the IGM clumping and ionization structure of the reionized gas.  In addition, Appendix \ref{app:convergence} of the current paper tests the effect of the domain structure on the mass and size distributions of optically thick absorbers.  We will discuss these quantities in detail in \S \ref{subsec:mass_sys}.  We adopt a power-law spectrum with specific intensity $J_{\nu} \propto \nu^{-1.5}$ between 1 and 4 Ry, in 5 frequency bins ($\nu$ is frequency).  This is motivated by stellar population synthesis models of metal poor populations \citep[e.g.][]{D'Aloisio_2019ApJ}.

\begin{figure*}
\centering
\resizebox{14cm}{!}{\includegraphics[trim={0cm 0cm 0cm 0cm},clip]{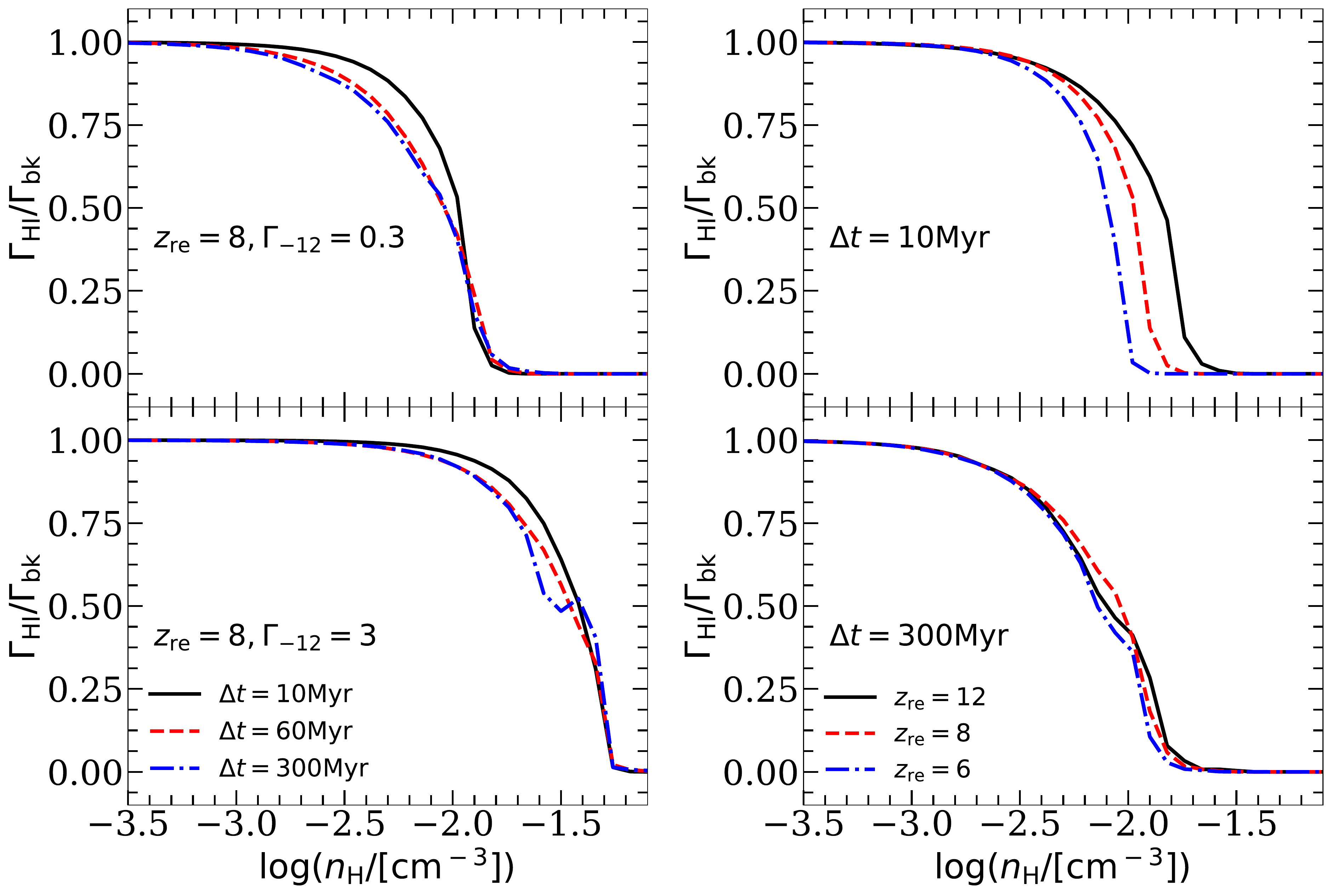}}
\vspace*{-2mm}
\caption{Evolution in the relationship between \HI\ photoionization rate ($\Gamma_{\rm HI}$) and proper hydrogen number density ($n_{\rm H}$) as the IGM responds to photoheating from reionization. The curves shows the median of $\Gamma_{\rm HI}$ normalized by the impinging background rate, $\Gamma_{\rm bk}$, in logarithmically spaced bins of $n_{\rm H}$. {\it Top Left:}  The effect of relaxation in the simulation with ($z_{\rm re}, \Gamma_{-12}, \delta/\sigma) = ( 8, 0.3, 0)$.  The black/solid, red/dashed, and blue/dot-dashed curves correspond to $z=7.9$, $7.5$, and $6.0$, respectively, $\Delta t=10$, $60$ and $300$ Myr after \zre. {\it Bottom Left:}  Same as top-left but with $\Gamma_{-12} = 3.0$. In the text we provide a simple model for how photoevaporation and relaxation change the $\Gamma_{\rm HI}$ vs. $n_{\rm H}$ relationship, in terms of density profiles and sizes of absorption systems.  {\it Top-Right:} The $\Gamma_{\rm HI}$ vs. $n_{\rm H}$ relationship at a fixed time interval, $\Delta t = 10$ Myr, from \zre$=12$ (black/solid), $8$ (red/dashed), and $6$ (blue/dot-dashed).  {\it Bottom-right:} Same but for $\Delta t = 300$ Myr.  The top-right panel shows differences that arise due to structure formation (see text). The bottom-right panel shows that these differences are largely erased by the relaxation/photoevaporative processes.   }
\label{fig:gamma_vs_nh}
\end{figure*}

Our small-scale simulations span a range of \zre\ and LyC intensities to sample the patchiness of the global reionization process (see Table 1 of Paper I). The impinging LyC intensity is parameterized by \GammaHI, the \HI photoionization rate in the source cells, expressed in units of $10^{-12}~{\rm s^{-1}}$. Our runs assume $\Gamma_{-12} = 0.3$ and $3.0$. The former is consistent with Ly$\alpha$ forest measurements \citep[e.g.][]{DAloisio2018} just after reionization ($z=5-6$), or perhaps during its tail end. The latter is intended to model the intensity near an over-density of sources.  The simulations also include a set of ``DC mode'' runs that model fluctuations away from the mean density on the box scale \citep{Gnedin2011}.   The box-scale density is parameterized by $\delta/\sigma$, the linearly extrapolated density contrast smoothed on the box scale, in units of its standard deviation.\footnote{Note that this ratio is independent of redshift.}  In addition to $\delta/\sigma =0$, i.e. cosmic mean density runs, we consider $\delta/\sigma = \pm \sqrt{3}$, which correspond to present-day linearly extrapolated over-densities of $\delta_0 = \pm 5.013$.\footnote{The values of $\delta/\sigma$ were chosen such that integrals over the Gaussian distribution of the box-scale density could be performed with three-point Gauss-Hermite Quadrature. See Paper I for more details.} 


\section{Photoionization rate vs density}
\label{sec:photoion}

We begin with the relationship between the photo-ionization rate and local hydrogen number density ($n_{\rm H}$), which characterizes self-shielding in our simulations.  Earlier works have studied this relationship in full cosmological simulations of reionization \citep{Rahmati2013,Chardin2018}. In this section, we examine how the hydrodynamic response of the IGM to photoheating drives evolution in the local relationship between $\Gamma_{\rm HI}$ and $n_{\rm H}$ during patchy reionization.  

In Figure~\ref{fig:gamma_vs_nh} we illustrate several key dependencies of the $\Gamma_{\rm HI}$ vs. $n_{\rm H}$ relationship.    The curves in the left column show the median $\Gamma_{\rm HI}$ vs. $n_{\rm H}$ at $\Delta t = 10$, 60, and 300 Myr from \zre.  The top and bottom panels correspond to $\Gamma_{-12} = 0.3$ and 3.0, respectively, both with $z_{\rm re} = 8$ and  $\delta/\sigma = 0$.  For simplicity, we focus on these two simulations but the trends described here hold more generally.  Consider the case with $\Gamma_{-12} = 0.3$ (top-left).  For all $\Delta t$, $\Gamma_{\rm HI}$ reaches 0 at approximately the same $\log(n_{\rm H}/{\rm cm}^{-3}) \approx -1.9$.  However, $\Gamma_{\rm HI}$ falls more steeply for times closer to \zre.  The bottom-left panel shows the same general trend.  A more intense ionizing background simply moves the self-shielding cutoff to higher densities.\footnote{The kinks at $\Gamma_{\rm HI}/\Gamma_{\rm bk} = 0.5$, seen especially in the bottom-left panel, arise from gas shadowed from ionizing radiation in one direction.}  

The right column of Figure~\ref{fig:gamma_vs_nh} shows how the relationship depends on \zre\ for fixed $\Gamma_{-12} = 0.3$. The different curves correspond to $z_{\rm re} = 12$, 8, and 6 at $\Delta t = 10$ (top) and $300$ (bottom) Myr after \zre.  Shortly after I-front passage, the decline in $\Gamma_{-12}$ is steeper at lower \zre. The bottom panel shows, however, that the $\Gamma_{\rm HI}$ vs. $n_{\rm H}$ curves approach a nearly identical form by $\Delta t = 300$ Myr.  As we will now discuss, all of these trends may be understood qualitatively in terms of the density profiles of the sinks, and how they are shaped by the competing effects of relaxation and structure formation.       

A toy model can provide insight into the results of Figure~\ref{fig:gamma_vs_nh}.  Consider a spherically symmetric absorber with a density profile, 

\begin{equation}
 n_{\rm H}(r) = 50~\bar{n}_{\rm H}(z) \left(\frac{r}{r_0}\right)^{-\alpha},
\end{equation}

\noindent where $\bar{n}_H(z)$ is the proper cosmic mean hydrogen number density, $r_0$ characterizes the size of the absorber, and $\alpha$ sets the steepness of the density profile.  We have normalized the profile such that $n_{\rm H} = 50 \bar{n}_{\rm H}$ at $r=r_0$.  We also take our fiducial $r_0$ to be $10h^{-1}$ kpc.  Both the over-density and $r_0$ are broadly motivated by the characteristic over-densities at which self-shielding occurs and the sizes of absorbers in our simulations (see \S 4.3 of Paper I, as well as \S \ref{sec:3d_sys} of the current paper).  However, the ensuing qualitative discussion does not depend on the particular choice of normalization.     We expose this absorber to an external isotropic ionizing background and solve the one-dimensional RT equation to obtain $\Gamma_{\rm HI}$ vs. $n_{\rm H}$ under the assumption that the gas is in photoionization equilibrium.\footnote{In detail, the photo-ionization rate towards the center of the absorber is
\begin{equation}
    \label{eq:gam_r}
    \Gamma_{\rm HI}(r) = \Gamma_{\rm bk} \exp(-\tau(r))
\end{equation}
where the opacity along the radial direction towards the center is given by
\begin{equation}
    \label{eq:tau_r}
    \tau(r) = \int_r^{\infty} dr' \sigma_{\rm HI} n_{\rm HI}(r')
\end{equation}
If the gas everywhere inside the absorber is in photo-ionizational equilibrium with the radiation, then we have
\begin{equation}
    \label{eq:photoeq}
    \Gamma_{\rm HI}(r) n_{\rm HI}(r) = \alpha_B(T) (1+\chi)(n_H(r) - n_{\rm HI}(r))^2
\end{equation}
These equations can be solved iteratively to obtain the profile of $\Gamma_{\rm HI}$ for an assumed density profile. For simplicity, we assume a uniform temperature of $T=10^4$ K.  }

\begin{figure}
\resizebox{8.5cm}{!}{  \includegraphics[]{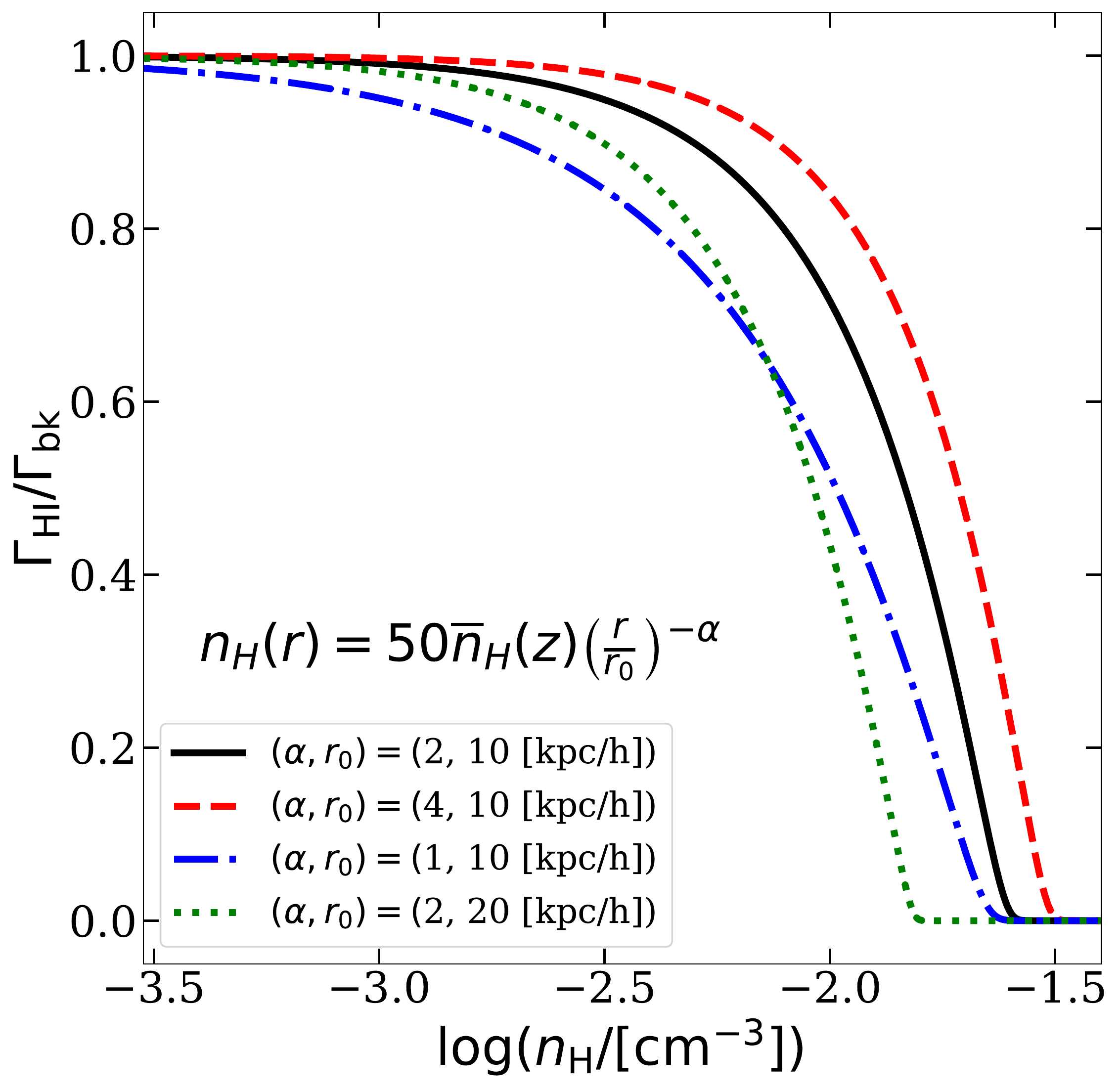}}
\vspace*{-5mm}
\caption{A toy model that we use to understand how photoevaporation and relaxation drive changes in the $\Gamma_{\rm HI} - n_{\rm H}$ relation. This model assumes a spherically-symmetric power-law density profile illuminated from the outside by an isotropic ionizing background.  The gas is assumed to be in photoionization equilibrium with the background.  We show results for $z=6$. Illustrative values for the power-law slope and absorber size are displayed. If photoevaporation makes the absorber density profiles shallower, and characteristically larger, for example, self-shielding will set in at lower $n_{\rm H}$. }
\label{fig:model_fig}
\end{figure}

 Figure~\ref{fig:model_fig} shows results from this model at $z=6$ for several combinations of $\alpha$ and $r_0$, where the latter is given in comoving units.  We have fixed the intensity of the impinging ionizing background such that the hydrogen photoionization rate far away from the absorber is $\Gamma_{\rm bk}= 3\times 10^{-13}$ s$^{-1}$.  The black curve with $(\alpha, r_0) = (2, 10\text{ kpc/h})$ corresponds to an isothermal density profile with size representative of the simulated systems that we will discuss in \S \ref{sec:3d_sys}. The other curves vary $\alpha$ and $r_0$ to mimic the effects of relaxation, photoevaporation, and structure formation.

 Increasing $r_0$ results in self-shielding setting in at smaller $n_{\rm H}$, i.e. $\Gamma_{\rm HI}$ falls off at smaller $n_{\rm H}$.  This occurs because a column density of $N_{\rm HI} = 1/\sigma_{912}$ is reached at smaller $n_H$ for larger values of $r_0$. (Here, $\sigma_{912}$ is the photoionization cross section of hydrogen at 1 Ry, or $\approx 912$ \AA.)   If the black/solid curve is a density peak at high redshift, the green/dotted curve represents that same peak at lower redshift after structure formation has increased its mass.  This provides a qualitative understanding of the trends seen in the top-right panel of Fig. \ref{fig:gamma_vs_nh}.  For lower \zre, structure formation has had more time to grow the sizes of the density peaks that serve as absorbers.  We thus see $\Gamma_{\rm HI}$ falling off at smaller $n_{\rm H}$.

 At fixed $r_0$, flattening the density profile (decreasing $\alpha$) also causes $\Gamma_{\rm HI}$ to fall off at smaller $n_{\rm H}$.  This effect, combined with possible evolution in $r_0$ during relaxation/photoevaporation, provides a simple picture for the evolution with $\Delta t$ seen in the left panels of Fig. \ref{fig:gamma_vs_nh}.  The relaxation and photoevaparation processes alter the density profiles of the absorbers, changing $\alpha$ and $r_0$ and driving the statistical evolution in Fig.\ref{fig:gamma_vs_nh}.  We note that these changes are complicated; they likely depend on the details of the initial configuration \citep{Shapiro2004}.
 
 Finally, while this single-absorber model provides insight into the trends seen in our simulations, we caution that the situation is more complicated in reality with a diverse population of absorbers.  For example, shortly after I-front passage much of the absorption may occur in diffuse gas within filaments (c.f. Figs. 3 and 4 in Paper I), in which case spherical symmetry is a poor approximation.  The model nonetheless provides some intuition for how both structure formation and the hydrodynamic response to photoheating drive evolution in the $\Gamma_{\rm HI}$ vs. $n_{\rm H}$ relation.     

\section{Contributions to the LyC Opacity}
\label{sec:LyC_opacity}

\begin{figure*}
\begin{center}
\resizebox{17cm}{!}{\includegraphics{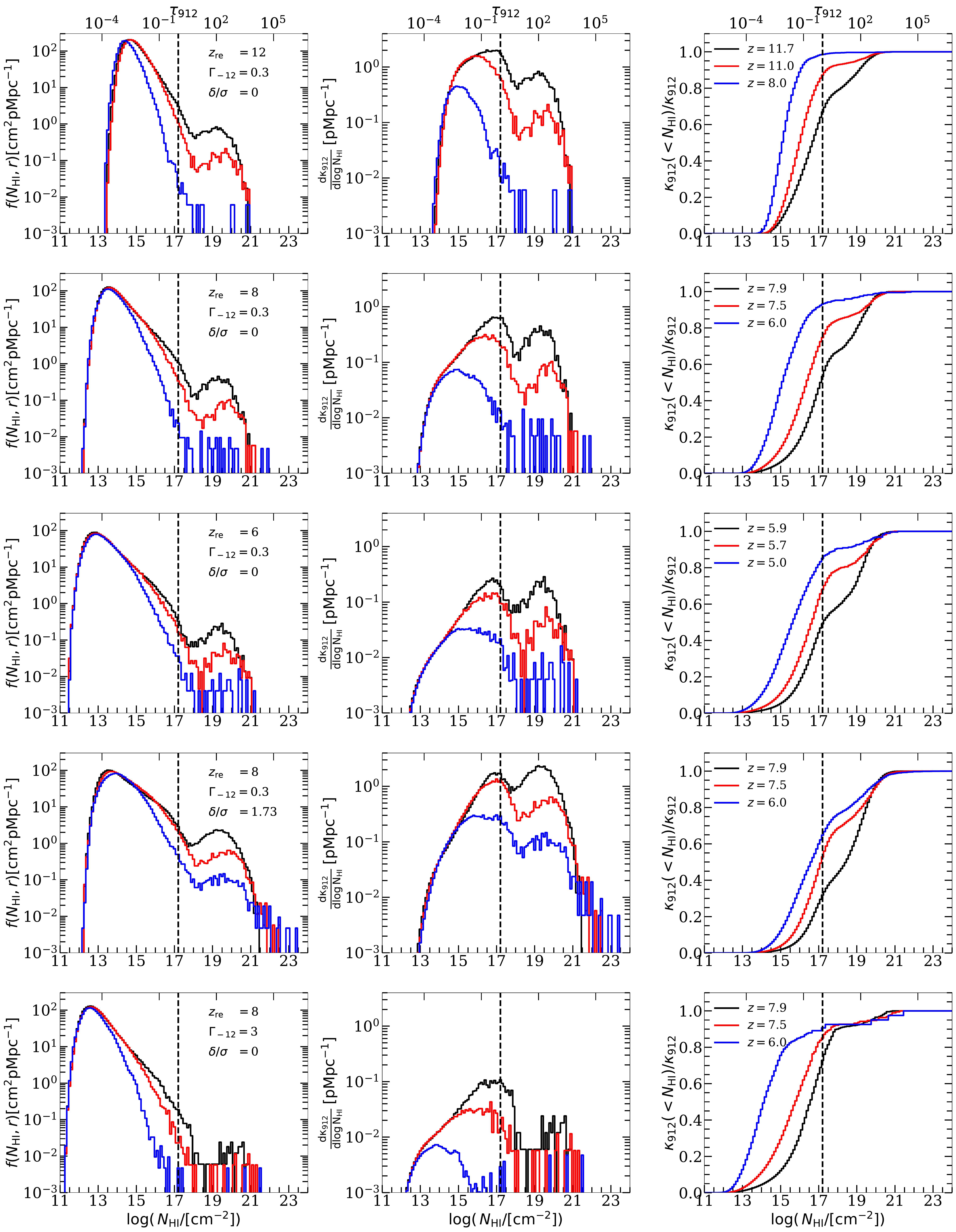}}
\end{center}
\vspace*{-3mm}
\caption{Contributions to the Lyman limit opacity of the IGM during reionization. {\it Left:} \HI\ column density distributions measured over 32\kpc-long skewers traced through our RT domains. 
The models are indicated in first panel of each row. The vertical black dashed lines indicate the $\taui=1$. The black, red, and blue curves correspond to $\Delta t = 10$, 60, and 300 \Myr after \zre, respectively. {\it Center:} Contribution to the absorption coefficient, $\kappa_{912}$, per logarithmic interval in $N_{\rm HI}$.  {\it Right:} Cumulative $\kappa_{\rm 912}(<N_{\rm HI})$ expressed as a fraction of the total $\kappa_{\rm 912}$. The fractional contribution to the opacity from columns above or below a given value can be read off of these panels. }
\label{fig:kappa}
\end{figure*}

Paper I explored the evolution of $\MFP$ in ionized gas during the relaxation process.   In this section, we aim to better understand what sets $\MFP$ by decomposing the Lyman limit opacity into its constitute absorption systems.

In Paper I, the MFP was calculated by tracing skewers of length $L_{\rm dom} = 32$ \kpc\ along each coordinate axis through the RT domains, and evaluating the outgoing flux $f_{\rm out} = e^{-\tau_{912}}$ along each skewer, where $\tau_{912} = \sigma_{912} N_{\rm HI}$.  Then the MFP was obtained using $\lambda_{912} = -L_{\rm dom}/\ln(\langle f_{\rm out} \rangle)$, where the average is taken over all segments.  It is straightforward to show that, in the limit $1 - \langle f_{\rm out} \rangle << 1$, the effective absorption coefficient in our simulation volume, $\kappa_{912} \equiv \lambda_{912}^{-1}$, can be expressed as

\begin{equation}
\kappa_{\rm 912} = \int^{N_{\rm max}}_{N_{\rm min}} f(N_{\rm HI}, r) (1-e^{-N_{\rm HI}\sigma_{\rm 912}}) dN_{\rm HI}
\label{eq:kappa}
\end{equation}
where the integral runs over all column densities and
\begin{equation}
\label{eq:fnh1r}
f(N_{\rm HI}, r) = \frac{\partial^2 N}{\partial N_{\rm HI} \partial r},
\end{equation}
is the number of segments per unit column, per unit length, i.e. the column density distribution. This conveniently re-casts the MFP calculation as an integral over column densities, allowing us to write the absorption coefficient per logarithmic interval in $N_{\rm HI}$ as \citep{Rahmati2018}
\begin{equation}
\begin{split}
\label{eq:dkappa}
 \frac{d\kappa_{\rm 912}}{d{\rm log}N_{\rm HI}}  = & \\ \frac{1}{\Delta {\rm log} N_{\rm HI}} & \int_{N_{\rm HI}}^{N_{\rm HI}+\Delta N_{\rm HI}} f(N_{\rm HI}, r) (1-e^{-N_{\rm HI}\sigma_{\rm 912}}) dN_{\rm HI},
\end{split}
\end{equation}
where we take $\Delta \log(N_{\rm HI}) = 0.1$.  Through eq.~(\ref{eq:dkappa}), we quantify the contribution to the opacity from different column densities, which will give us insight into the properties of the gaseous structures responsible for setting the MFP.  Equation (\ref{eq:kappa}) neglects spatial correlations between the absorption systems; it is identical to the expression for $\kappa_{912}$ in the widely used model of Poisson distributed absorbers \citep{1980ApJ...240..387P}. Recognizing that this expression can be derived in the appropriate limit from our definition of \mfp\ in Paper I, we find that the former is quite accurate in practice. 

In Figure~\ref{fig:kappa} we show, from left to right, the column density distribution (Eq.~\ref{eq:fnh1r}), the absorption coefficient per $\Delta \log(N_{\rm HI})$ (Eq.~\ref{eq:dkappa}), and the cumulative fractional contribution to the absorption coefficient (Eq.~\ref{eq:kappa}), as functions of $N_{\rm HI}$, for the 32 kpc/h domain segments from our simulations.  The simulation parameters are indicated in the left-most panel in each row.  
The black, red, and blue curves correspond to $\Delta t = 10$, 60, and 300 Myr after $z_{\rm re}$. (For the $z_{\rm re} = 6$ case, the blue corresponds to $\Delta t = 244$ Myr, since we did not run the simulation past $z=5$.)  In each panel, the top axis shows $\tau_{912}$ and the vertical dashed lines denote $\tau_{912} = 1$.

Some key qualitative features are evident for all the simulations shown in Figure~\ref{fig:kappa}.   There is a distinct peak around $N_{\rm HI} = 10^{19}$ cm$^{-2}$ in the column density distribution of recently ionized gas (black curves) in all but the $\Gamma_{-12} = 3.0$ case.  This peak disappears by $\Delta t = 300$ Myr except in the over-dense run ($\delta/\sigma = \sqrt{3}$), although it is suppressed significantly there as well.  The steady decrease in the abundance of segments with columns $\gtrsim 10^{15}$ cm$^{-2}$ reflects the hydrodynamic response of the gas after ionization.  At early times, there is abundant small-scale structure with $\tau_{912}\sim 1$ but it is erased over $\Delta t \sim 100$ Myr by Jeans smoothing and photoevaporation.   

The second column shows that $d \kappa_{912}/ d \log N_{\rm HI}$ exhibits two prominent peaks shortly after I-front passage ($\Delta t = 10$ Myr)   -- the first at $N_{\rm HI} \sim 10^{17}$ cm$^{-2}$ and the second at $N_{\rm HI} \sim 10^{19}$ cm$^{-2}$.  As we will discuss below, these peaks represent two physically distinct types of optically thick absorbers.   The relative heights of the peaks -- and therefore their relative contributions -- are set by the interplay between all three simulation parameters. Generally, lower $\Gamma_{-12}$, lower $z_{\rm re}$, and larger environmental density ($\delta/\sigma$) all increase the contribution of the $N_{\rm HI}\sim 10^{19}$ cm$^{-2}$ peak.  However, as the gas relaxes, this peak all but disappears except in the over-dense run. Most $N_{\rm HI}\gtrsim 10^{17}$ cm$^{-2}$ systems do not survive the relaxation/evaporation except where the local over-density is large.\footnote{More specifically, by ``local over-density" we mean $\delta/\sigma$, where $\delta$ is smoothed over our $1h^{-1}$ Mpc box scale.}

From the right-most panels, we can glean the fractional contribution of optically thick systems to the $\kappa_{912}$ of our boxes.  The contributions vary significantly with the environmental parameters. For example, in the second panel with $(z_{\rm re}, \Gamma_{-12}, \delta/\sigma) = (8,0.3,0)$, the $\tau_{912}>1$ systems account for $ 47\%$ of the opacity at a time $\Delta t = 10$ Myr after I-front passage.  At the same $\Delta t$, they account for only $27\%$ when $\Gamma_{-12}$ is a factor of 10 higher (bottom panel). In the over-dense run (fourth panel), they account for $68 \%$.  Notably, $\tau_{912} >1$ systems still account for $ 35\%$ of the opacity in this run when $\Delta t = 300$ Myr. This is in contrast to all other runs, for which the contribution from $\tau_{912} >1$ systems becomes $\leq 15~\%$ in the relaxed limit.

By the time a few hundred Myr have passed since \zre, the {\it local} MFP in an average $\sim 1$ \mpc\ patch of the IGM is set by low column-density gas. For example, in our simulation with $(z_{\rm re}, \Gamma_{-12}, \delta/ \sigma) = (8, 0.3, 0)$,  approximately $80 \%$ of the opacity comes from sight lines with $N_{\rm HI} < 10^{16}$ cm$^{-2}$ when $\Delta t = 300$ Myr.  This owes to almost all of the high-column systems being wiped out by the relaxation process. Although the {\it local} MFP in this relaxed $1 h^{-1}$ Mpc patch is set mainly by optically thin gas, the global MFP becomes regulated by optically thick absorbers in over-dense patches, recently reionized patches, or by neutral gas that has yet to be reionized.  Indeed, the fourth row of Fig. \ref{fig:kappa} illustrates that optically thick absorbers have greater longevity in over-dense regions. We conclude by noting that cosmic expansion and an increasing ionizing background intensity ($\Gamma_{-12}$) both act to increase the contribution of LLSs to $\kappa_{912}$ well after reionization.  Future work should explore how the picture presented here transitions to the well-studied opacity structure of the $z=2-4$ IGM.

\section{Demographics of Optically Thick Absorbers During reionization}
\label{sec:demographics}

\begin{figure*}
    \begin{center}
\resizebox{8.5cm}{!}{\includegraphics[trim={.2cm 0cm 0cm 0cm},clip]{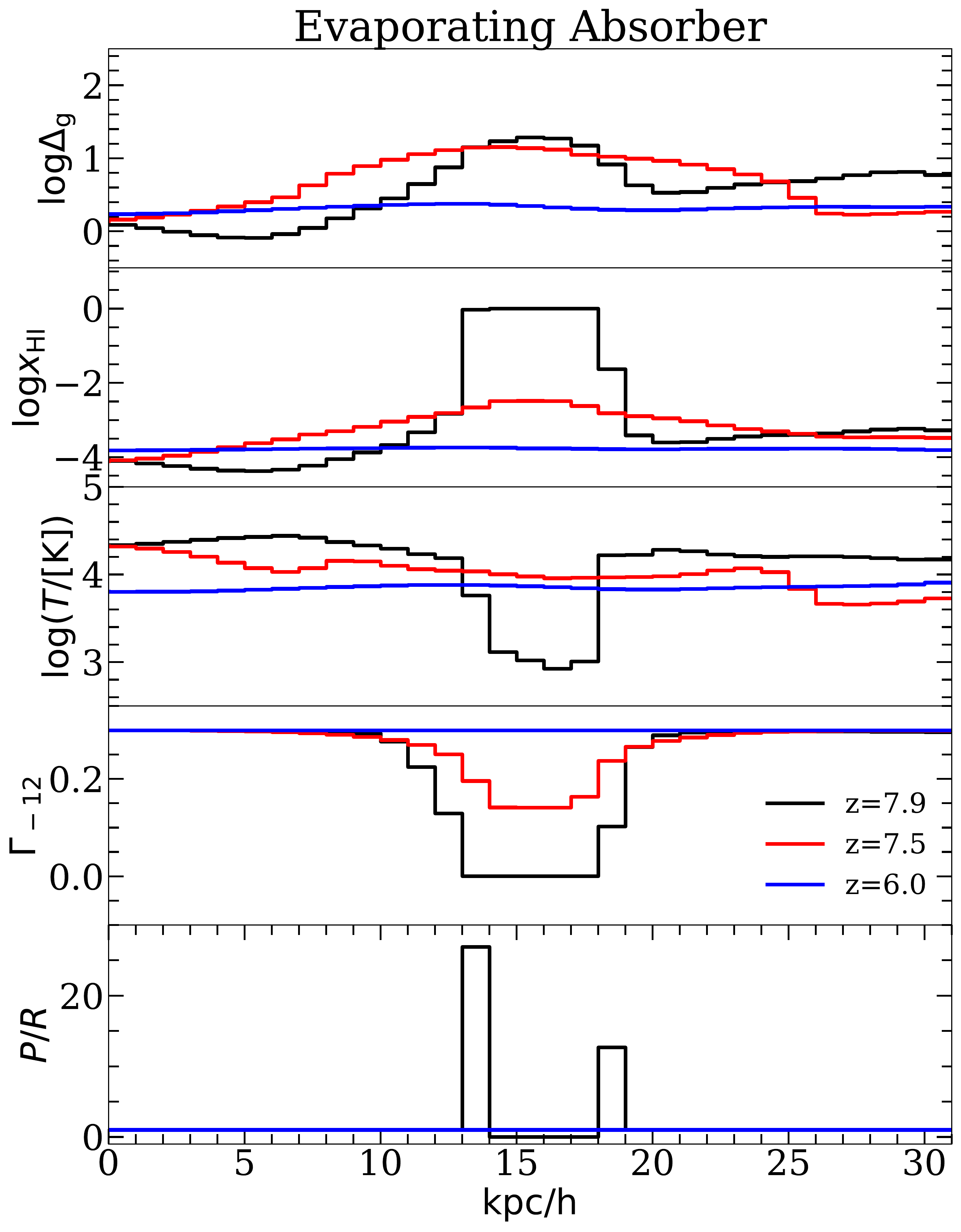}}
\resizebox{8.5cm}{!}{\includegraphics[trim={0cm 0cm .2cm 0cm},clip]{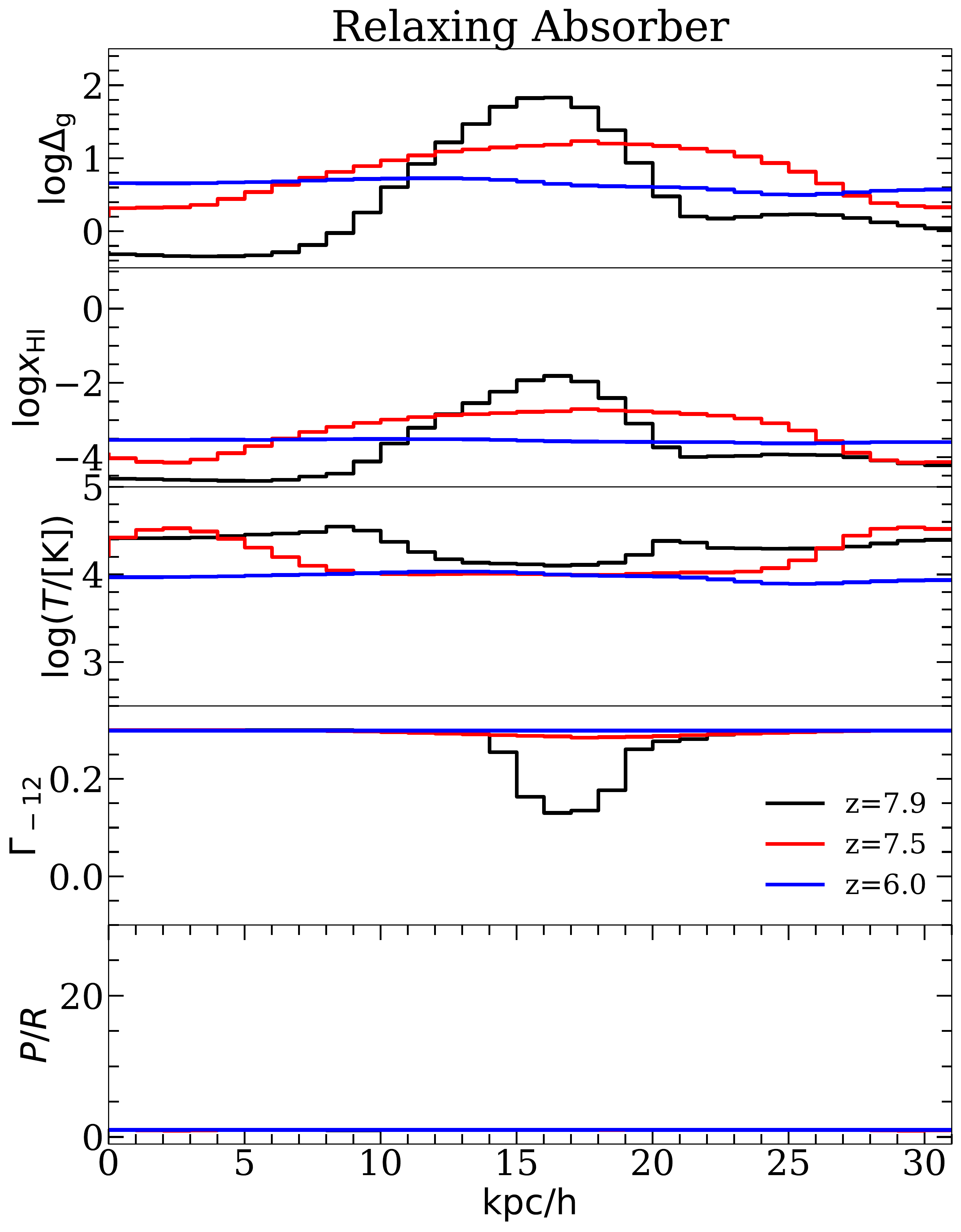}}
\vspace*{-2mm}
\caption{Example skewers through one of our simulations illustrating two distinct types of optically thick absorption systems.  These examples are from the simulation with $(z_{\rm re}, \Gamma_{-12}, \delta/\sigma) = (8, 0.3, 0)$.  From top to bottom, the rows show the gas over-density, \HI\ fraction, temperature, \HI\ photoionization rate, and the ratio of photoionization rate to recombination rate. The last quantity is unity for gas in photoionization equilibrium.  Each panel shows the time evolution of the sight line with the black, red, and blue lines corresponding to $\Delta t = 10$, $60$,and $300$ Myr after \zre, respectively. The sight line on the left has column densities of $\log N_{\rm HI}/[\mathrm{cm}^{-2}] = 18.8$, $16.3$, and $14.7$ at $\Delta t = 10$, $60$, and $300$ Myr, respectively.  On the right, we have $\log N_{\rm HI}/[\mathrm{cm}^{-2}] = 17.3$, $16.3$, and $15.2$, respectively.  At $\Delta t = 10$ Myr, we classify the absorber on the left as ``evaporating," characterized by its neutral core and the presence of gas out of photoionization equilibrium. We call the absorber on the right ``relaxing."  This sight line is highly ionized, yet optically thick at $\Delta t = 10$ Myr.  Unlike the case on the left, the I-front likely never transitioned to D-type as it swept past this sight line. }
\label{fig:skewer_example}
\end{center}
\end{figure*}

\begin{figure}
\resizebox{8.5cm}{!}{  \includegraphics[]{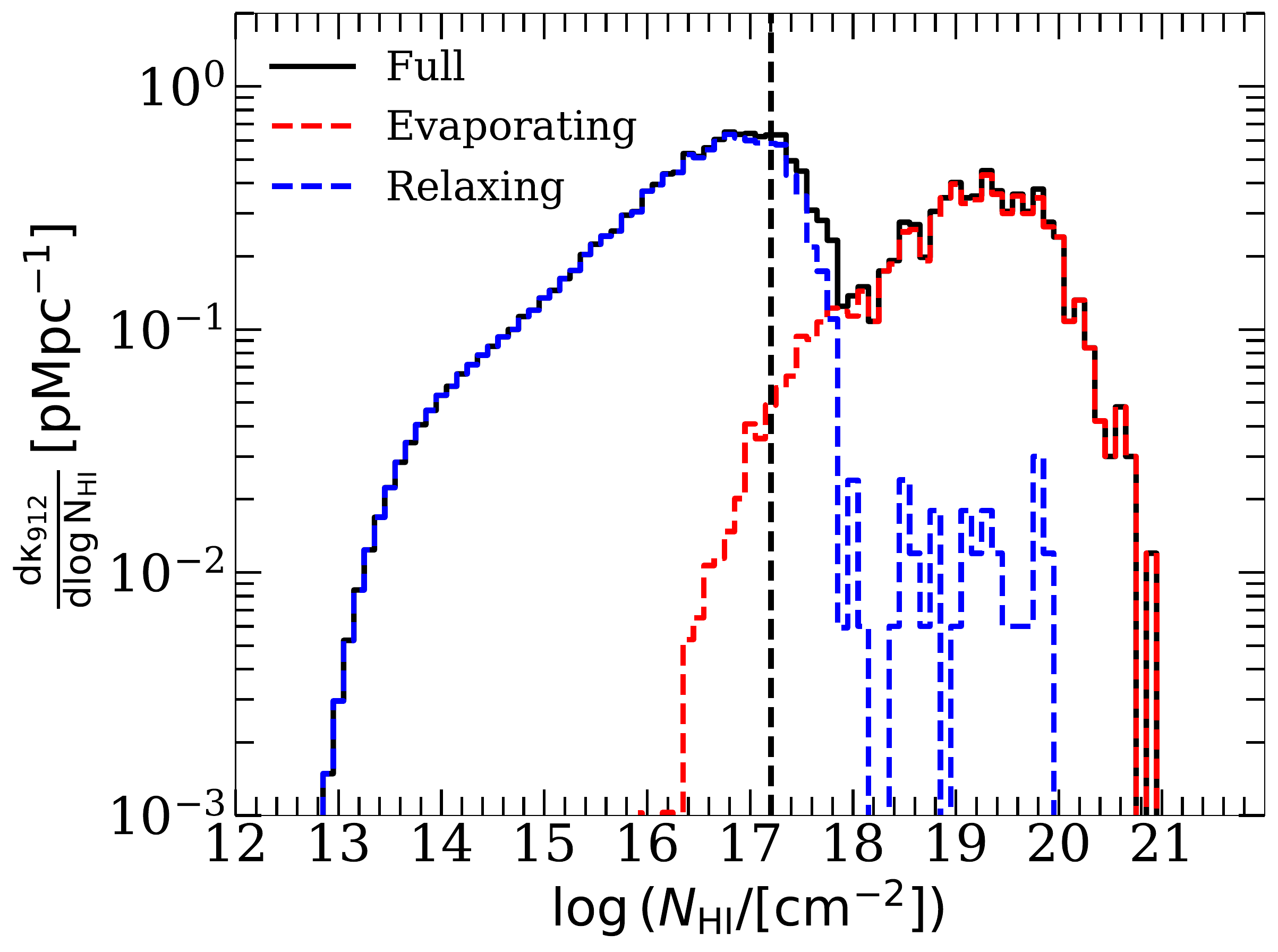}}
\vspace*{-5mm}
\caption{The contribution of segments containing evaporating (red/dashed) and relaxing (blue/dashed) absorbers to the differential opacity, ${\rm d}\kappa_{\rm 912}/{\rm log}\NHI$, shown as solid black curve. The distributions are from the $(\Gamma_{-12}, z_{\rm re}, \delta/\sigma) = (0.3, 8, 0)$ simulation. The optically thick absorbers consist of two distinct populations of sinks: (1) density peaks with fully neutral cores (evaporating); (2) highly ionized absorbers in photo-ionization equilibrium with the ionizing background (relaxing).    }
\label{fig:bumps}
\end{figure}

Figure \ref{fig:kappa} shows that the distributions of column densities in our simulations are bi-modal throughout much of the relaxation process.  Here we illustrate that the bi-modality arises from two distinct types of absorption systems in recently reionized gas.

\begin{figure*}\center
\resizebox{18.5cm}{!}{\includegraphics[trim={4cm 2cm 3cm 0cm},clip]{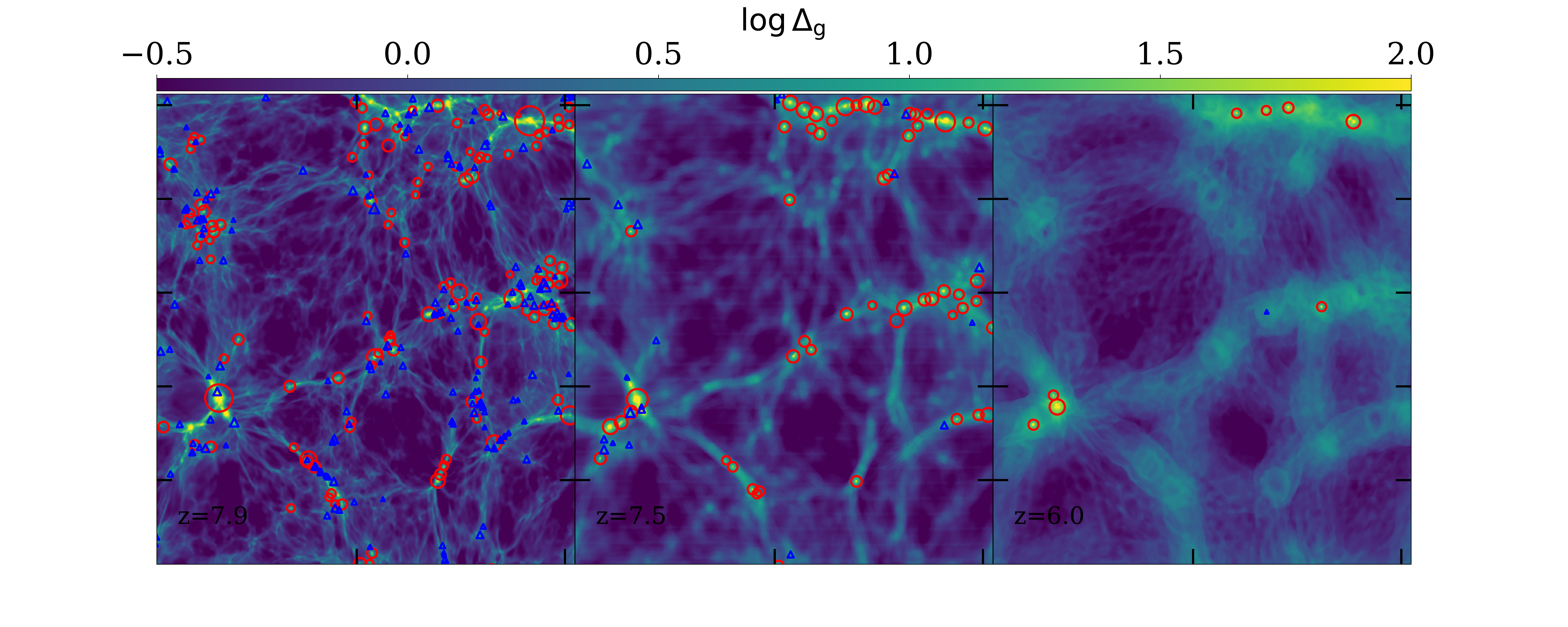}}
\vspace*{-3mm}
\caption{ Visualization of absorbers identified in our simulation with  $( z_{\rm re}, \Gamma_{-12}, \delta/\sigma) = (8, 0.3, 0)$.  See main text for details on our identification procedure. Gas over-density is shown in color scale and each slice is 32\kpc\ thick. The red circles and blue triangles show evaporating and relaxing systems, respectively. The size of each symbol is proportional to the effective diameter of the corresponding absorber. 
The same slice is shown at $z=7.9$ (left), $7.5$ (center) and $6.0$ (right), corresponding to $\Delta t = 10, 60$, and $300$ Myr. The surviving systems (right panel) are located at the highest density peaks and contain fully neutral cores, hence their classification as evaporating. }
\label{fig:slice}
\end{figure*}

The sight lines with the highest column densities, which comprise the peaks located around $\log N_{\rm HI}/[\rm{cm}^{-2}] \approx 19$,  almost always contain absorbers with neutral fractions close to unity.   Note that these high-$N_{\rm HI}$ peaks disappear over a time scale of $\sim 100$ Myr because the absorbers are photo-evaporated by the ionizing background impinging on their outskirts.  To illustrate this, the left panels of Figure \ref{fig:skewer_example} show an example sight line intersecting an absorption system undergoing photo-evaporation.  The skewer is drawn from one of the RT domains in the simulation with $(z_{\rm re}, \Gamma_{-12}, \delta/\sigma) = (8, 0.3, 0)$. From top to bottom, we show the gas density ($\Delta_g$; in units of the cosmic mean), the \HI\ fraction ($x_{\rm HI}$), the gas temperature ($T$), the photo-ionization rate, and the ratio of the photo-ionization rate to the recombination rate, $P/R \equiv \Gamma_{\rm HI}/(\mathcal{R}/n_{\rm HI})$, where $\mathcal{R} = \alpha_B(T)n_e n_{\rm HII}$, and $\alpha_B$ is the case B recombination coefficient of hydrogen.  The black, red, and blue curves correspond to snapshots in time at $\Delta t = 10, 60$, and $300$ Myr after \zre, respectively. The total column density of the sight line evolves from $\log N_{\rm HI}/[\rm{cm}^{-2}] = 18.8$ at $\Delta t = 10$ Myr, to $16.3$ and $14.7$ at $\Delta t = 60$ and $300$ Myr, respectively. 
At $\Delta t = 10$ Myr, the absorber located near the center of the sight line is peaked in density. (Though, the sight line does not necessarily pass through the center of the peak in three dimensions.)  The neutral fraction reaches unity near the center, where $\Gamma_{-12}$ dips to zero -- the result of self-shielding. And the temperature of the peak is $\sim 1,000$ K, much lower than the photo-ionized gas outside of it.     A key feature of this system is that the gas at the boundaries of the neutral core is out of photo-ionization equilibrium, as evidenced by the spikes in $P/R$ shown in the bottom panel.  By $\Delta t = 60$ Myr, the density peak has been smoothed out considerably.  The gas is highly ionized and has reached photo-ionization equilibrium, with the equilibration time scale being $\sim 1/\Gamma_{\rm HI} \sim 10^{13}~\rm{s} = 3\times 10^5$ years at the center of the peak (see 3rd row).   The gas is evacuated from the system by $\Delta t = 300$ Myr and the absorber is gone.\footnote{We have checked the example sight lines of Fig. \ref{fig:skewer_example} in an adiabatic version of the simulation.  We have verified that the density peaks are still present at $z=6$, indicating that their peculiar motions cannot account for the observed flattening.}  In what follows, we we will refer to systems like these, which exhibit large deviations in $P/R$ from unity, as ``evaporating systems."     

The situation is considerably different in the right panels. The column densities of this sight line are $\log N_{\rm HI}/[\rm{cm}^{-2}] = 17.3$, $16.3$, and $15.2$ at $\Delta t = 10$, $60$ and $300$ Myr, respectively.    Despite the sight line having an opacity of $\tau_{912}\approx 1.2$ at $\Delta t = 10$ Myr, there is never any fully neutral core associated with the central density peak. In fact, the peak is highly ionized and in photo-ionization equilibrium at all snapshots shown. However, this peak, too, is erased by Jeans pressure smoothing within $\Delta = 300$ Myr. We will refer to this type of system as a ``relaxing" absorber.  We note that the key distinction between these two classifications is the timescale over which I-fronts are able to ionize all of the gas in the system.  Evaporating systems are those in which the I-fronts get ``stuck" as they climb the steep density gradient, transitioning from R-type to D-type \citep[see e.g.][]{Shapiro2004}. In contrast, relaxing systems may have been quickly ionized as I-fronts swept supersonically through the region, perhaps never transitioning to D-type.

To support the picture that the absorbers consist of two distinct populations, Figure \ref{fig:bumps} shows what happens when we select sight lines based on whether the gas is in photo-ionization equilibrium. The black/solid curve shows $d \kappa_{912}/d \log N_{\rm HI}$ at $\Delta t = 10$ Myr from the full sample of sight lines in our simulation with $(\Gamma_{-12}, z_{\rm re}, \delta/\sigma) = (0.3, 8, 0)$.  As a proxy for evaporating systems, the red/dashed curve corresponds to sight lines with $| P/R | \geq 1.5$ anywhere along the sight line.\footnote{Note that this is not a perfect selection criterion, as RT geometry can sometimes result in misidentifying evaporating systems.} The blue/dashed curve corresponds to those with $|P/R| < 1.5$.  That the two peaks separate cleanly suggests that the high-column peaks in Fig. \ref{fig:kappa} are comprised almost entirely of evaporating systems with fully neutral gas.  Although we show an illustrative example here, we find that this separation occurs for all the simulations in our suite.   In the next section, we will explore in more detail the physical properties of these systems.

\section{Properties of Absorption Systems} \label{sec:3d_sys}

\begin{figure*}
\begin{center}
\resizebox{16.0cm}{!}{\includegraphics{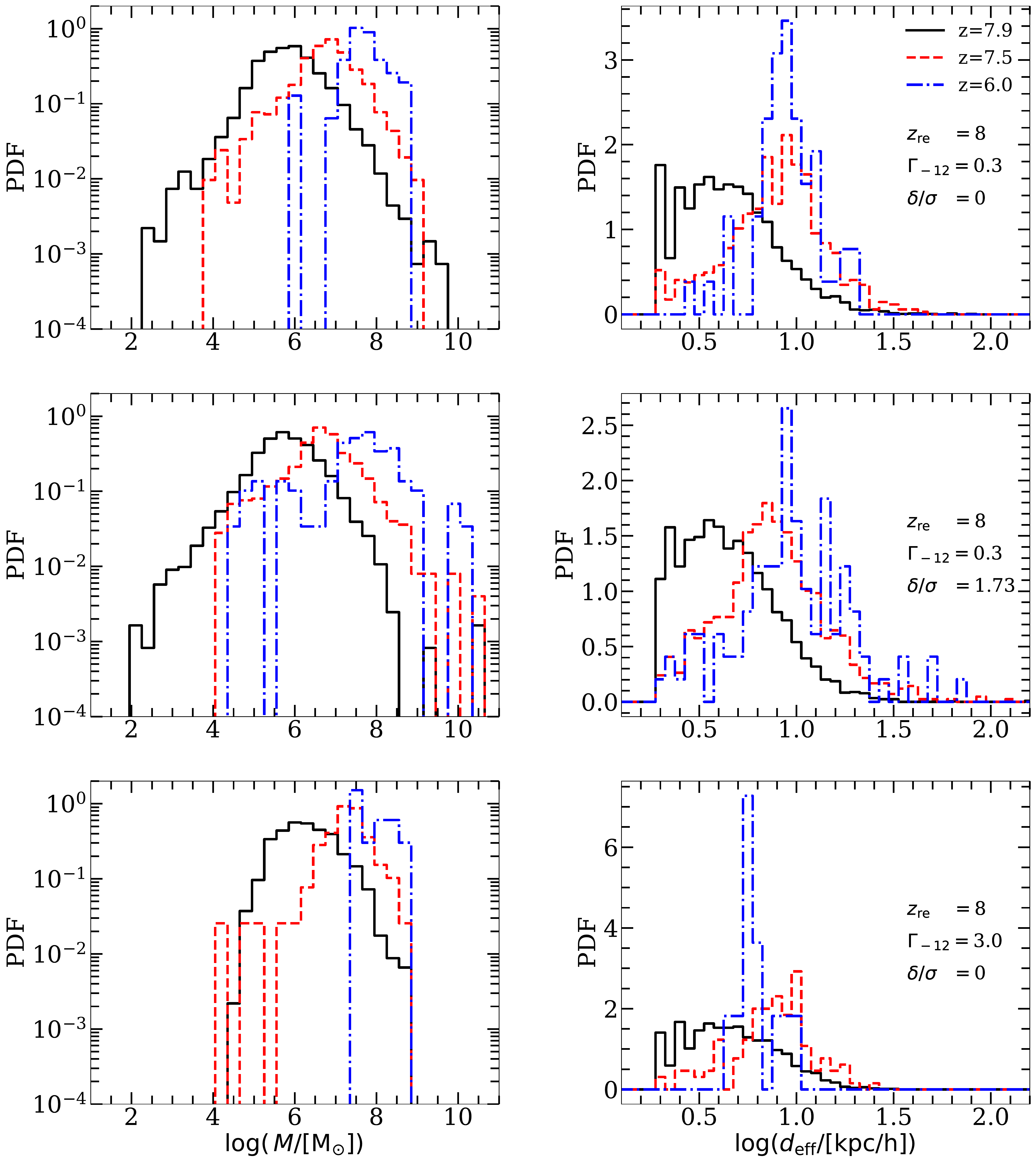}}
\end{center}
\caption{Distribution of absorber masses (left) and effective diameters (right) in three of our simulations. Each row corresponds to a different simulation with parameters provided in the legends of the right panels.   The black/solid, red/dashed, and blue/dot-dashed curves correspond to $\Delta t = 10$ ,$60$ and $300$ Myr, respectively. The effective diameter is $d_{\rm eff} = 2{(3V/4\pi)}^{1/3}$, where $V$ is absorber volume. The absorption systems responsible for setting the LyC opacity during reionization can be as small as the Jeans scale of the cold, pre-reionization gas. In a cold-dark-matter dominated universe, the smallest absorbers may be beyond the resolution limits of our simulations.   }
\label{fig:mass_vol}
\end{figure*}

Here we explore the properties of the 3-dimensional structures responsible for the 1-dimensional absorbers studied above.  Accomplishing this requires a prescription for identifying self-shielding structures.  Following \citet{Rahmati2018}, we define a self-shielding structure (or ``absorber") to be a simply-connected group of cells with neutral fraction above a threshold $x_{\rm HI}^{\rm thresh}$.   We use a variable threshold (which depends on $\Gamma_{-12}$, \zre, $\delta/\sigma$) chosen to optimally link the population of $\tau_{912} \geq 1$ absorbers to the 3D systems that create them.\footnote{This differs from \citet{Rahmati2018}; they used a constant $x_{\rm HI}^{\rm thresh}=0.01$.}  Appendix \ref{app:thresh} describes our procedure for calculating $x_{\rm HI}^{\rm thresh}$.  Essentially, we optimize $x_{\rm HI}^{\rm thresh}$ to maximize completeness of our sample, while also minimizing the number of optically thin 1-dimensional sight lines through our absorbers.  We then form simply connected groups out of the cells above $x_{\rm HI}^{\rm thresh}$.  Table \ref{tab:xHIthresh} lists $x_{\rm HI}^{\rm thresh}$ for our simulation parameters.  Our values are typically $x_{\rm HI}^{\rm thresh}\approx 0.005$, but can be as low as $x_{\rm HI}^{\rm thresh}\approx 0.001$ for $\Gamma_{-12} = 3$ or $\delta/\sigma = -\sqrt{3}$. 

To illustrate our identification procedure, Figure~\ref{fig:slice} shows example slices through the density field from our run with $(\Gamma_{-12}, z_{\rm re}, \delta/\sigma) = (0.3, 8, 0)$.  From left to right, the panels exhibit the effects of relaxation on the density structure. Self-shielding structures are marked by red circles or blue triangles, with the sizes of the symbols proportional to the effective diameters, $d_{\rm eff}\equiv 2(3 V/4 \pi)^{1/3}$, where $V$ is volume. Red circles correspond to evaporating systems (with $|P/R| \geq 1.5$ somewhere in the system) and blue triangles correspond to relaxing systems ($|P/R| < 1.5$).  Visual inspection of Fig \ref{fig:slice} reveals that the self-shielding structures trace out the filaments and tend to be clustered around the nodes.  By $\Delta t = 300$ Myr (right panel) the few systems that remain are at the highest density peaks and are classified as evaporating because they possess fully neutral cores.

\subsection{Masses \& Sizes}\label{subsec:mass_sys}

Figure~\ref{fig:mass_vol} shows the time-evolution of absorber mass (left column) and size (right column) in three of our simulations, where the former includes both dark matter and gas mass.  To obtain dark matter masses, we smooth particles onto the hydro grid using cloud-in-cell interpolation. We characterize the sizes of the systems by their effective diameters, $d_{\rm eff}$.  Note, however, that many are not spherically symmetric, especially early in the relaxation process.   

The typical mass and size of an absorption system shifts from smaller to larger values during the relaxation process.  In our runs with $\Gamma_{-12} = 0.3$ (top two rows), the masses and sizes range from $M\sim 10^{4} - 10^8$ $M_{\odot}$ and $d_{\rm eff} \sim \mathrm{a~few} - 20 h^{-1}$ kpc at $\Delta t = 10$ Myr.  By $\Delta t = 300$ Myr, most systems have $M\sim 10^7 - 10^{9}$ $M_{\odot}$ and $d_{\rm eff} \sim 5 - 30 h^{-1}$ kpc.  The shift to larger systems with time is straightforward to understand.  Just after I-front passage, the gas contains a large population of small self-shielding systems, but they evaporate/relax quickly. The larger systems survive for longer time scales.   The middle row shows results from one of our over-dense simulations with $\delta/\sigma = \sqrt{3}$. At $\Delta t = 10$ Myr ($z=7.9$), the distribution of absorber masses appears similar to the case with $\delta/\sigma = 0$. Note, however, that the local density enhancement results in a couple of very massive $M\sim 10^{10} - 10^{11} h^{-1}$ M$_\odot$ systems. Indeed, although not visible in the middle-right panel, these rare systems have $d_{\rm eff} \sim 150 h^{-1}$ kpc and are in the extreme tail of the distribution.     By $\Delta t = 300$ Myr, the absorbers in over-dense box span a broader range of masses, with the high-mass tail of the distribution extending up to $> 10^{10}~h^{-1}$ M$_\odot$. The density enhancement also results in significantly more surviving absorbers. For example, at $\Delta t = 10$ Myr, we find $4,533 (4,069)$ absorbers in the run with $\delta /\sigma = 0(\sqrt{3})$.  By $\Delta t = 300$ Myr, the over-dense run has 98 surviving absorbers, a factor of $\approx 2$ more than the 52 found in the mean-density box. Turning to the bottom row, higher $\Gamma_{-12}$ generally eliminates the systems with lower mass.  This occurs because the smaller, milder over-densities are not able to self-shield against the intense background and are erased on a quicker timescale.  By $\Delta t = 300$ Myr, the surviving systems are somewhat smaller than the case with $\Gamma_{-12}=0.3$ because self-shielding occurs at higher densities.  We find $1,519 (11)$ absorbers in the $\Gamma_{-12}=3.0$ run at $\Delta t = 10(300)$ Myr. 

We quantify the ratio of gas mass to total mass for the absorption systems shown in Figure~\ref{fig:mass_vol}. From top to bottom, the mean fractions are 0.10, 0.12 and 0.092, respectively, at $\Delta t = 10$ Myr. Note that the simulation with $\delta/\sigma = \sqrt{3}$ has a slightly higher mean fraction compared to the  $\delta/\sigma = 0$ case. On the other hand, the run with enhanced ionizing background, $\Gamma_{-12}=3.0$, exhibits a reduced fraction. 
Most of the systems are below the cosmic baryon fraction, $\frac{\Omega_{\rm b}}{\Omega_{\rm m}-\Omega_{\rm b}}=0.187$. At  $\Delta t = 60$ Myr, the fractions drop to 0.075, 0.091 and 0.072. By $\Delta t = 300$ Myr, there are fewer systems and a large scatter in gas fraction.  The mean values actually increase to 0.083, 0.145 and 0.187, respectively, owing the a small number of systems with large ($\approx 0.2-0.5$) gas fractions.

One concern about the results of Fig. \ref{fig:mass_vol} is that the distribution of absorber sizes might be affected by the domain structure of our RT. For example, one could imagine that a source plane along the boundary of a domain might intersect an absorber, causing its inner regions to be artificially ionized and splitting the absorber into two.  We tested explicitly for such effects.  In Appendix \ref{app:convergence} we show that we obtain similar absorber size distributions even when using larger domains. Of greater concern, however, is the fact that our simulations do not capture the smallest absorption systems that exist in an adiabatically cooling IGM before I-front passage. Based on the numerical convergence tests also presented in Appendix \ref{app:convergence}, we caution that we may be over-estimating the typical mass of absorption systems by a factor of $\approx 2$, and the typical sizes by $\sim 25\%$. In a cold-dark-matter dominated universe, the smallest absorbers may even be beyond the resolution limits of our simulations

\subsection{Over-densities}

\begin{figure}\center
\resizebox{8.cm}{!}{  \includegraphics[]{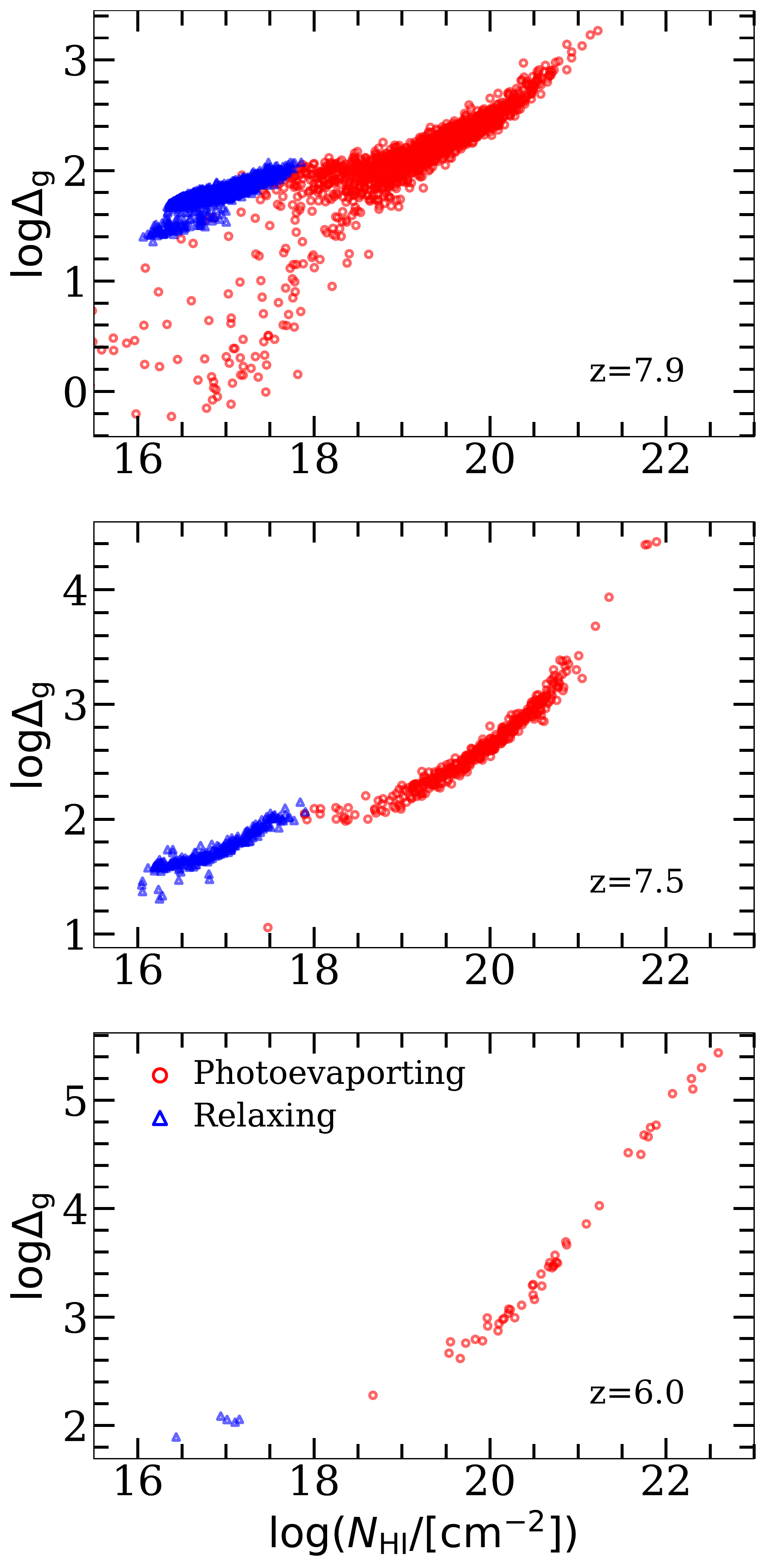}}
\vspace*{-3mm}
\caption{Relationship between $n_{\rm HI}$-weighted mean over-density ($\Delta_g$) and column density ($N_{\rm HI}$) for self-shielding absorption systems.  Absorbers are identified as simply connected regions above a neutral fraction threshold, as described in the main text.  The red circle and blue triangles correspond to evaporating and relaxing absorbers, respectively.  We show results from our simulation with $(z_{\rm re}, \Gamma_{-12}, \delta/ \sigma) = (8, 0.3, 0)$.  From top to bottom, the panels correspond to $\Delta t = 10$, $60$, and $300$ Myr after $z_{\mathrm{re}}$.  The hydrodynamic response of the gas to reionization depletes the total number of absorbers and raises the characteristic density above which self-shielding occurs.   }
\label{fig:NHI_and_nH}
\end{figure}

Figure \ref{fig:NHI_and_nH} shows scatter plots of the $n_{\rm HI}$-weighted mean gas over-density of the absorbers vs. their column densities. To calculate both quantities, we integrate over randomly oriented skewers traced through the peaks of $n_{\rm HI}$ in the absorbers. We average over the skewers to obtain a single value for each absorber.  As in Fig. \ref{fig:slice}, the red circles and blue triangles correspond to evaporating and relaxing absorbers, respectively. For brevity, we show results for our fiducial simulation with $(z_{\rm re}, \Gamma_{-12}, \delta/ \sigma) = (8, 0.3, 0)$, but the results are qualitatively similar in our other simulations.  

The progress of time from top to bottom illustrates how relaxation depletes the abundance of absorbers that exist in the unrelaxed IGM following I-front passage. At all snapshots, $\Delta_g$ increases with $N_{\rm HI}$ except for a flattening that occurs around $\log(N_{\rm HI}/[\mathrm{cm}^{-2}]) \sim 17-19$.  This is qualitatively similar to the self-shielding ``plateau" between $\log(N_{\rm HI}/[\mathrm{cm}^{-2}]) \sim 18-20$ seen in the simulations of \citet{McQuinn2011b}. At $\Delta t = 10$ Myr (top panel), self-shielding sets in at the characteristic over-density of $\Delta_g \sim 25$.  However, by $\Delta t = 300$ Myr (bottom), almost all of the absorbers have $\Delta_g \gtrsim 100$ and $\log(N_{\rm HI}/[\mathrm{cm}^{-2}]) \gtrsim 19.5$.  Most also possess neutral cores as indicated by their classification as evaporating systems. Although not shown in Fig. \ref{fig:NHI_and_nH}, larger $\Gamma_{-12}$ values shift the scatter plots upwards in $\Delta_g$ and further deplete the absorber population.  Larger box-scale densities, $\delta/\sigma$, have the opposite effect.  

The unrelaxed absorbers at $\Delta t = 10$ Myr exhibit a higher degree of scatter compared to later times.  Notably, there are absorbers that deviate from the main $\Delta_g$ vs $N_{\rm HI}$ relation.  In fact, there even appear to be under-dense absorbers. These systems result from the complete or partial shadowing of the LyC radiation by nearby over-densities. For example, a region in the shadow of a large density peak may remain neutral until the peak is sufficiently smoothed for the I-front to break through.  Although these occurrences are already rare in our simulations, we caution that our 2-direction RT may exaggerate their prevalence. More realistically, radiation could come from many directions as nearby sources turn on during reionization.

\subsection{Ionization and Thermal states}
\label{subsec:gas_prop}

\begin{figure*}
\center
\resizebox{18cm}{!}{\includegraphics{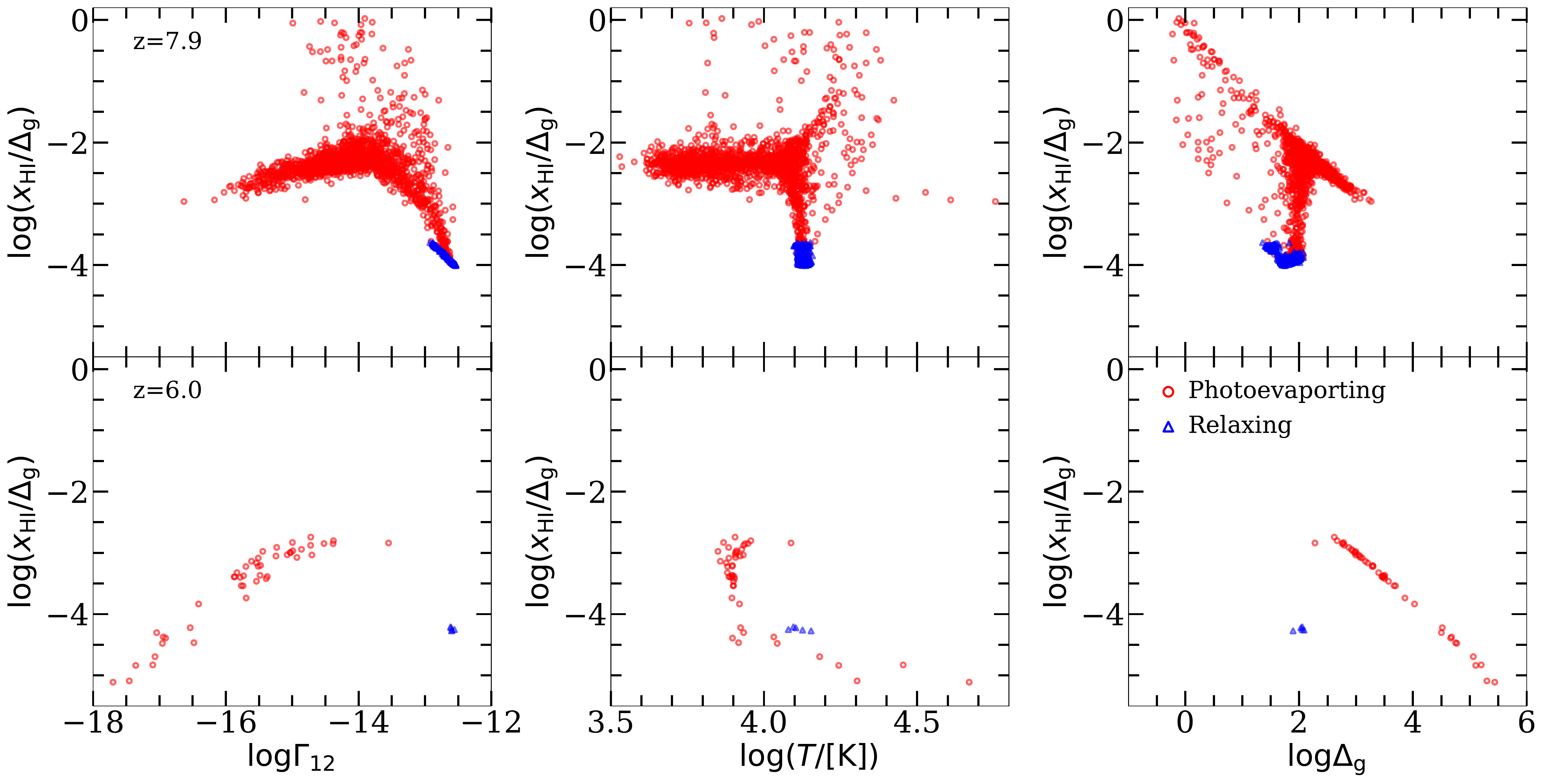}}
\vspace*{-5mm}
\caption{ Ionization and thermal states of self-shielding structures identified in our simulation with $(z_{\rm re}, \Gamma_{-12}, \delta/\sigma) = (8, 0.3, 0)$. The top and bottom rows correspond to two snapshots in time, $z=7.9$ and $6$, respectively. The evaporating and systems are shown as red circles and relaxing systems are shown as blue triangles. From left to right, we show the $n_{\rm HI}$-weighted mean \HI\ photoionization rate, temperature, and gas over-density of absorbers, vs. $x_{\rm HI}/\Delta_g$.  These quantities are calculated by tracing randomly oriented skewers passing through the peak $n_{\rm HI}$ in each absorber.  Relaxing systems are relegated to a relatively small region of the parameter space, consistent with highly ionized gas in equilibrium with the ionizing background.   }
\label{fig:ion_thermal}
\end{figure*}

An informative way to characterize the state of absorbers in our simulations is through the photo-ionization equilibrium condition in highly ionized gas,
\begin{equation}
    \label{eq:equil}
    \frac{x_{\rm HI}}{\Delta_g}  \approx \frac{\alpha_B(T) \langle n_{\rm H} \rangle (1+\chi)   }{\Gamma_{\rm HI}}
\end{equation}
where $\chi = n_{\rm He}/n_{\rm H} \approx 1.08$ accounts for singly ionized He, $\alpha_B$ is the case B recombination rate of H, and $\langle n_H \rangle$ is the cosmic mean H number density.  Note that, because of our RT domain setup, $\Gamma_{\rm HI}$ is approximately uniform throughout highly ionized gas in our simulations.  The utility of characterizing the absorbers with equation (\ref{eq:equil}) is that, for gas in equilibrium, the right-hand side should depend only on $T$, to a good approximation.  To the extent that recently ionized gas has a relatively narrow range of temperatures, we therefore expect ${x_{\rm HI}}/{\Delta_g}$ to take on a narrow range of values for relaxing systems.  This should not be true for evaporating systems, however.  

In Fig. \ref{fig:ion_thermal} we show $x_{\rm HI}/\Delta_g$ versus $\Gamma_{\rm HI}$, $T$, and $\Delta_g$, for our identified absorbers.  All quantities are $n_{\rm HI}$-averaged.\footnote{As in the last section, we obtain mean values obtained by averaging over randomly oriented skewers through the peaks of $n_{\rm HI}$.}  For brevity the figure contains results only from the simulation with $(\Gamma_{-12}, z_{\rm re}, \delta/\sigma)=(0.3,8,0)$, but the same basic trends hold in all of our simulations.  The top and bottom panels correspond to $\Delta t = 10$ and $300$ Myr after $z_{\rm re}$, respectively.   The evaporating systems (red circles) show a large range of values, whereas the relaxing systems (blue triangles) reside in a smaller region of the parameter space, as we had anticipated above. 

In the left panels, the relaxing systems mostly have mean $\Gamma_{\rm HI}$ near the background value of $\Gamma_{\rm HI} = 3\times 10^{-13}$ s$^{-1}$. In contrast, the evaporating systems span a range of $\Gamma_{\rm HI}$; at $z=6$ they extend down to $\Gamma_{\rm HI} = 10^{-18}$ s$^{-1}$.  This is consistent with the evaporating systems possessing neutral cores. The trend of $x_{\rm HI}/\Delta_g$ increasing with $\Gamma_{\rm HI}$ for evaporating systems reflects that lower densities are less self-shielded against the background.  In the top panel, the scatter towards $x_{\rm HI}/\Delta_g \sim 1$ owes to mildly over-dense gas shadowed from the background by nearby density peaks.  

In the middle column, the evaporating systems span a range of temperatures, from $\sim 4,000 - 32,000$ K.  The lower temperatures are similar to virial temperatures of halos with $M\sim 10^6$ $M_{\odot}$.  The higher temperatures may result from very recently ionized, low-density gas \citep{D'Aloisio_2019ApJ} or from the collapse of dense systems.  In contrast, the relaxing systems are all clustered around temperatures characteristic of photoionized gas in equilibrium with the background, $T\sim 14,000$ K. In all panels at $z=7.9$ ($\Delta t = 10$ Myr), there are transitional systems bridging the evaporating and relaxing populations.  The populations become more separated at late times.   

In the right columns, we see that the relaxing systems, with characteristic over-densities $\Delta_g \sim 100$, reside at the lower end of the $\Delta_g$-distribution.  The few evaporating systems with $\Delta_g$ significantly less than $100$ correspond to diffuse neutral gas in the shadows of nearby peaks. A large fraction of evaporating systems exhibit a tight correlation with $\Delta_g$.  This owes to the fact that $x_{\rm HI}/\Delta_g \sim 1/\Delta_g$ for largely neutral systems.  By $z = 6.0$, the surviving optically thick absorbers are highly over-dense; almost all are fully neutral systems centered on halos.   The analysis presented in the three preceding sub-sections broadly supports our earlier discussion of two distinct populations of optically thick absorbers.


\section{Conclusions}\label{sec:conclusions}

We have studied the properties of ionizing photon sinks during reionization.  Our study was based on the suite of high-resolution radiative hydrodynamics simulations of Paper I, which model the self-shielding and hydrodynamic response of the IGM in the wake of I-fronts.   Our main findings can be summarized as follows:

\begin{itemize}

\item The density and ionization structures of the IGM evolve considerably after $z_{\rm re}$ as the gas relaxes and density peaks are photo-evaporated.   We have quantified these effects with the $\Gamma_{\rm HI}$ vs. $n_{\rm H}$ relationship, which is often used to model self-shielding in cosmological simulations without RT \citep{Rahmati2013, Chardin2018}.   The evolution can be understood qualitatively with a simple model that considers the competing effects of gravitational growth and density-peak smoothing by the ionizing background.

\item We measured the column-density distributions in our simulations to characterize the absorption systems that contribute most to the LyC opacity. These distributions depend on environmental parameters: $z_{\rm re}$, $\Gamma_{-12}$, and $\delta/\sigma$.  Broadly speaking, systems with $N_{\rm HI} \geq 10^{17.2}$ cm$^{-2}$ ($\tau_{912} \geq 1$) account for $\sim 50\%$ of the opacity within $\Delta t \sim 10$ Myr of $z_{\rm re}$.  The onus shifts to lower columns during relaxation, with $\tau_{912} \geq 1$ systems contributing just $\lesssim 10 \%$ by $\Delta t = 300$ Myr.  Higher environmental density and lower $\Gamma_{-12}$ generally increases the contribution of  $\tau_{912}\geq 1$ systems.  For example, in our over-dense run with $\delta /\sigma = \sqrt{3}$, the $\tau_{912}\geq 1$ systems still account for $35\%$ of the opacity at $\Delta t = 300$ Myr. 

\item
For $\sim 50$ Myr after $z_{\rm re}$, the $\tau_{912} \geq 1$ absorbers are comprised of two distinct populations; (1) $\Delta_g \gtrsim 100$ systems with fully neutral cores; (2) Highly ionized systems at milder over-densities, which are in photo-ionization equilibrium with the ionizing background.  The latter are just dense enough to be self-shielding but were not dense enough at $z_{\rm re}$ to halt the passing I-front.  The relative abundances of these two populations depends on $z_{\rm re}$, $\Gamma_{-12}$, and $\delta/\sigma$.  After a couple hundred Myr, relaxation and photo-evaporation erase the vast majority of these structures.  The surviving structures have $\Delta \gtrsim 500$, fully neutral cores, and are clustered preferentially in large-scale over-densities.  
 
 \item
 The characteristic masses and sizes of $\tau_{912} \geq 1$ absorbers evolve from smaller to larger during the relaxation process.  Within $\sim 50$ Myr after $z_{\rm re}$, total masses and sizes are typically in the range $M=10^{4} - 10^{8}$ M$_{\odot}$ and $d_{\rm eff} = \mathrm{a~few} - 20 h^{-1}$ kpc. However, by $\Delta t \sim 300$ Myr, they are $M=10^{7} - 10^{9}$ M$_{\odot}$ and $d_{\rm eff} = 5 - 30 h^{-1}$ kpc, respectively.  Again, these properties depend on the environmental parameters $z_{\rm re}$, $\Gamma_{-12}$, and $\delta/\sigma$. 
 
 \end{itemize}
 
 Finally, we note some implications for the small characteristic masses and sizes of sinks in a cold-dark-matter universe. First, these scales highlight the extreme computational challenge of modelling the sinks in cosmological simulations of reionization, especially during the first $\sim 50$ Myr of the relaxation process.    This consideration is particularly relevant for modeling the high-$z$ IGM if reionization ended around $z=5$, with a significant fraction of the gas still relaxing at those times \citep[e.g.][]{Kulkarni2019,Keating2020,Nasir2020}. Second, it is worth noting that the broad conclusions presented here could be considerably different in other dark matter cosmologies with a damping scale in the primordial power spectrum, e.g. warm dark matter, fuzzy dark matter, and some varieties of self-interacting dark matter. In this sense, measurements of the IGM opacity near reionization can, in principle, test dark matter models at scales and densities that are not probed by other methods.

\vspace{1cm}
We thank Hy Trac for providing his RadHydro code, and George Becker and Fred Davies for helpful comments on this manuscript. AD's group was supported by HST grant HST-AR15013.005-A, NASA grant 19-ATP19-0191, and NSF grant 2045600.  Computations were performed with NSF XSEDE allocations TG-AST120066 and TG-PHY210041.

\bibliography{bibliography}{}

\begin{thebibliography}{}
\expandafter\ifx\csname natexlab\endcsname\relax\def\natexlab#1{#1}\fi
\providecommand{\url}[1]{\href{#1}{#1}}
\providecommand{\dodoi}[1]{doi:~\href{http://doi.org/#1}{\nolinkurl{#1}}}
\providecommand{\doeprint}[1]{\href{http://ascl.net/#1}{\nolinkurl{http://ascl.net/#1}}}
\providecommand{\doarXiv}[1]{\href{https://arxiv.org/abs/#1}{\nolinkurl{https://arxiv.org/abs/#1}}}

\bibitem[{{Altay} {et~al.}(2013){Altay}, {Theuns}, {Schaye}, {Booth}, \& {Dalla
  Vecchia}}]{Altay2013}
{Altay}, G., {Theuns}, T., {Schaye}, J., {Booth}, C.~M., \& {Dalla Vecchia}, C.
  2013, \mnras, 436, 2689, \dodoi{10.1093/mnras/stt1765}

\bibitem[{{Altay} {et~al.}(2011){Altay}, {Theuns}, {Schaye}, {Crighton}, \&
  {Dalla Vecchia}}]{Altay2011}
{Altay}, G., {Theuns}, T., {Schaye}, J., {Crighton}, N.~H.~M., \& {Dalla
  Vecchia}, C. 2011, \apjl, 737, L37, \dodoi{10.1088/2041-8205/737/2/L37}

\bibitem[{{Becker} {et~al.}(2021){Becker}, {D'Aloisio}, {Christenson}, {Zhu},
  {Worseck}, \& {Bolton}}]{2021arXiv210316610B}
{Becker}, G.~D., {D'Aloisio}, A., {Christenson}, H.~M., {et~al.} 2021, arXiv
  e-prints, arXiv:2103.16610.
\newblock \doarXiv{2103.16610}

\bibitem[{{Bouwens} {et~al.}(2015){Bouwens}, {Illingworth}, {Oesch}, {Trenti},
  {Labb{\'e}}, {Bradley}, {Carollo}, {van Dokkum}, {Gonzalez}, {Holwerda},
  {Franx}, {Spitler}, {Smit}, \& {Magee}}]{Bowens2015}
{Bouwens}, R.~J., {Illingworth}, G.~D., {Oesch}, P.~A., {et~al.} 2015, \apj,
  803, 34, \dodoi{10.1088/0004-637X/803/1/34}

\bibitem[{{Bouwens} {et~al.}(2021){Bouwens}, {Oesch}, {Stefanon},
  {Illingworth}, {Labbe}, {Reddy}, {Atek}, {Montes}, {Naidu}, {Nanayakkara},
  {Nelson}, \& {Wilkins}}]{2021arXiv210207775B}
{Bouwens}, R.~J., {Oesch}, P.~A., {Stefanon}, M., {et~al.} 2021, arXiv
  e-prints, arXiv:2102.07775.
\newblock \doarXiv{2102.07775}

\bibitem[{{Cain} {et~al.}(2021){Cain}, {D'Aloisio}, {Gangolli}, \&
  {Becker}}]{2021arXiv210510511C}
{Cain}, C., {D'Aloisio}, A., {Gangolli}, N., \& {Becker}, G.~D. 2021, arXiv
  e-prints, arXiv:2105.10511.
\newblock \doarXiv{2105.10511}

\bibitem[{{Chardin} {et~al.}(2018){Chardin}, {Kulkarni}, \&
  {Haehnelt}}]{Chardin2018}
{Chardin}, J., {Kulkarni}, G., \& {Haehnelt}, M.~G. 2018, \mnras, 478, 1065,
  \dodoi{10.1093/mnras/sty992}

\bibitem[{{Crighton} {et~al.}(2019){Crighton}, {Prochaska}, {Murphy},
  {O'Meara}, {Worseck}, \& {Smith}}]{Crighton2019}
{Crighton}, N. H.~M., {Prochaska}, J.~X., {Murphy}, M.~T., {et~al.} 2019,
  \mnras, 482, 1456, \dodoi{10.1093/mnras/sty2762}

\bibitem[{{D'Aloisio} {et~al.}(2018){D'Aloisio}, {McQuinn}, {Davies}, \&
  {Furlanetto}}]{DAloisio2018}
{D'Aloisio}, A., {McQuinn}, M., {Davies}, F.~B., \& {Furlanetto}, S.~R. 2018,
  \mnras, 473, 560, \dodoi{10.1093/mnras/stx2341}

\bibitem[{{D'Aloisio} {et~al.}(2019){D'Aloisio}, {McQuinn}, {Maupin}, {Davies},
  {Trac}, {Fuller}, \& {Upton Sanderbeck}}]{D'Aloisio_2019ApJ}
{D'Aloisio}, A., {McQuinn}, M., {Maupin}, O., {et~al.} 2019, \apj, 874, 154,
  \dodoi{10.3847/1538-4357/ab0d83}

\bibitem[{{D'Aloisio} {et~al.}(2020){D'Aloisio}, {McQuinn}, {Trac}, {Cain}, \&
  {Mesinger}}]{Daloisio2020}
{D'Aloisio}, A., {McQuinn}, M., {Trac}, H., {Cain}, C., \& {Mesinger}, A. 2020,
  \apj, 898, 149, \dodoi{10.3847/1538-4357/ab9f2f}

\bibitem[{{Davies} {et~al.}(2021){Davies}, {Bosman}, {Furlanetto}, {Becker}, \&
  {D'Aloisio}}]{2021arXiv210510518D}
{Davies}, F.~B., {Bosman}, S. E.~I., {Furlanetto}, S.~R., {Becker}, G.~D., \&
  {D'Aloisio}, A. 2021, arXiv e-prints, arXiv:2105.10518.
\newblock \doarXiv{2105.10518}

\bibitem[{{Finkelstein} {et~al.}(2015){Finkelstein}, {Ryan}, {Papovich},
  {Dickinson}, {Song}, {Somerville}, {Ferguson}, {Salmon}, {Giavalisco},
  {Koekemoer}, {Ashby}, {Behroozi}, {Castellano}, {Dunlop}, {Faber}, {Fazio},
  {Fontana}, {Grogin}, {Hathi}, {Jaacks}, {Kocevski}, {Livermore}, {McLure},
  {Merlin}, {Mobasher}, {Newman}, {Rafelski}, {Tilvi}, \&
  {Willner}}]{2015ApJ...810...71F}
{Finkelstein}, S.~L., {Ryan}, Russell~E., J., {Papovich}, C., {et~al.} 2015,
  \apj, 810, 71, \dodoi{10.1088/0004-637X/810/1/71}

\bibitem[{{Finkelstein} {et~al.}(2019){Finkelstein}, {D'Aloisio},
  {Paardekooper}, {Ryan}, {Behroozi}, {Finlator}, {Livermore}, {Upton
  Sanderbeck}, {Dalla Vecchia}, \& {Khochfar}}]{2019ApJ...879...36F}
{Finkelstein}, S.~L., {D'Aloisio}, A., {Paardekooper}, J.-P., {et~al.} 2019,
  \apj, 879, 36, \dodoi{10.3847/1538-4357/ab1ea8}

\bibitem[{{Furlanetto} \& {Oh}(2005)}]{Furlanetto_2005MNRAS}
{Furlanetto}, S.~R., \& {Oh}, S.~P. 2005, \mnras, 363, 1031,
  \dodoi{10.1111/j.1365-2966.2005.09505.x}

\bibitem[{{Gnedin} {et~al.}(2011){Gnedin}, {Kravtsov}, \& {Rudd}}]{Gnedin2011}
{Gnedin}, N.~Y., {Kravtsov}, A.~V., \& {Rudd}, D.~H. 2011, \apjs, 194, 46,
  \dodoi{10.1088/0067-0049/194/2/46}

\bibitem[{{Iliev} {et~al.}(2005{\natexlab{a}}){Iliev}, {Scannapieco}, \&
  {Shapiro}}]{2005ApJ...624..491I}
{Iliev}, I.~T., {Scannapieco}, E., \& {Shapiro}, P.~R. 2005{\natexlab{a}},
  \apj, 624, 491, \dodoi{10.1086/429083}

\bibitem[{{Iliev} {et~al.}(2005{\natexlab{b}}){Iliev}, {Scannapieco}, \&
  {Shapiro}}]{Iliev2005}
---. 2005{\natexlab{b}}, \apj, 624, 491, \dodoi{10.1086/429083}

\bibitem[{{Keating} {et~al.}(2020){Keating}, {Weinberger}, {Kulkarni},
  {Haehnelt}, {Chardin}, \& {Aubert}}]{Keating2020}
{Keating}, L.~C., {Weinberger}, L.~H., {Kulkarni}, G., {et~al.} 2020, \mnras,
  491, 1736, \dodoi{10.1093/mnras/stz3083}

\bibitem[{{Kim} {et~al.}(2013){Kim}, {Partl}, {Carswell}, \&
  {M{\"u}ller}}]{Kim2013}
{Kim}, T.-S., {Partl}, A.~M., {Carswell}, R.~F., \& {M{\"u}ller}, V. 2013,
  \aap, 552, A77, \dodoi{10.1051/0004-6361/201220042}

\bibitem[{{Kulkarni} {et~al.}(2019){Kulkarni}, {Keating}, {Haehnelt}, {Bosman},
  {Puchwein}, {Chardin}, \& {Aubert}}]{Kulkarni2019}
{Kulkarni}, G., {Keating}, L.~C., {Haehnelt}, M.~G., {et~al.} 2019, \mnras,
  485, L24, \dodoi{10.1093/mnrasl/slz025}

\bibitem[{{Mao} {et~al.}(2019){Mao}, {Koda}, {Shapiro}, {Iliev}, {Mellema},
  {Park}, {Ahn}, \& {Bianco}}]{Mao2019}
{Mao}, Y., {Koda}, J., {Shapiro}, P.~R., {et~al.} 2019, arXiv e-prints,
  arXiv:1906.02476.
\newblock \doarXiv{1906.02476}

\bibitem[{{McGreer} {et~al.}(2015){McGreer}, {Mesinger}, \&
  {D'Odorico}}]{McGreer2015}
{McGreer}, I.~D., {Mesinger}, A., \& {D'Odorico}, V. 2015, \mnras, 447, 499,
  \dodoi{10.1093/mnras/stu2449}

\bibitem[{{McQuinn} {et~al.}(2007){McQuinn}, {Lidz}, {Zahn}, {Dutta},
  {Hernquist}, \& {Zaldarriaga}}]{McQuinn2007}
{McQuinn}, M., {Lidz}, A., {Zahn}, O., {et~al.} 2007, \mnras, 377, 1043,
  \dodoi{10.1111/j.1365-2966.2007.11489.x}

\bibitem[{{McQuinn} {et~al.}(2011){McQuinn}, {Oh}, \&
  {Faucher-Gigu{\`e}re}}]{McQuinn2011b}
{McQuinn}, M., {Oh}, S.~P., \& {Faucher-Gigu{\`e}re}, C.-A. 2011, \apj, 743,
  82, \dodoi{10.1088/0004-637X/743/1/82}

\bibitem[{{Miralda-Escud{\'e}} {et~al.}(2000){Miralda-Escud{\'e}}, {Haehnelt},
  \& {Rees}}]{Miralda-Escude_2000ApJ}
{Miralda-Escud{\'e}}, J., {Haehnelt}, M., \& {Rees}, M.~J. 2000, \apj, 530, 1,
  \dodoi{10.1086/308330}

\bibitem[{{Nasir} \& {D'Aloisio}(2020)}]{Nasir2020}
{Nasir}, F., \& {D'Aloisio}, A. 2020, \mnras, 494, 3080,
  \dodoi{10.1093/mnras/staa894}

\bibitem[{{Paresce} {et~al.}(1980){Paresce}, {McKee}, \&
  {Bowyer}}]{1980ApJ...240..387P}
{Paresce}, F., {McKee}, C.~F., \& {Bowyer}, S. 1980, \apj, 240, 387,
  \dodoi{10.1086/158244}

\bibitem[{{Park} {et~al.}(2016){Park}, {Shapiro}, {Choi}, {Yoshida}, {Hirano},
  \& {Ahn}}]{Park2016}
{Park}, H., {Shapiro}, P.~R., {Choi}, J.-h., {et~al.} 2016, ArXiv e-prints.
\newblock \doarXiv{1602.06472}

\bibitem[{{Planck Collaboration} {et~al.}(2018){Planck Collaboration},
  {Aghanim}, {Akrami}, {Ashdown}, {Aumont}, {Baccigalupi}, {Ballardini},
  {Banday}, {Barreiro}, {Bartolo}, {Basak}, {Battye}, {Benabed}, {Bernard},
  {Bersanelli}, {Bielewicz}, {Bock}, {Bond}, {Borrill}, {Bouchet}, {Boulanger},
  {Bucher}, {Burigana}, {Butler}, {Calabrese}, {Cardoso}, {Carron},
  {Challinor}, {Chiang}, {Chluba}, {Colombo}, {Combet}, {Contreras}, {Crill},
  {Cuttaia}, {de Bernardis}, {de Zotti}, {Delabrouille}, {Delouis}, {Di
  Valentino}, {Diego}, {Dor{\'e}}, {Douspis}, {Ducout}, {Dupac}, {Dusini},
  {Efstathiou}, {Elsner}, {En{\ss}lin}, {Eriksen}, {Fantaye}, {Farhang},
  {Fergusson}, {Fernandez-Cobos}, {Finelli}, {Forastieri}, {Frailis},
  {Franceschi}, {Frolov}, {Galeotta}, {Galli}, {Ganga}, {G{\'e}nova-Santos},
  {Gerbino}, {Ghosh}, {Gonz{\'a}lez-Nuevo}, {G{\'o}rski}, {Gratton},
  {Gruppuso}, {Gudmundsson}, {Hamann}, {Hand ley}, {Herranz}, {Hivon}, {Huang},
  {Jaffe}, {Jones}, {Karakci}, {Keih{\"a}nen}, {Keskitalo}, {Kiiveri}, {Kim},
  {Kisner}, {Knox}, {Krachmalnicoff}, {Kunz}, {Kurki-Suonio}, {Lagache},
  {Lamarre}, {Lasenby}, {Lattanzi}, {Lawrence}, {Le Jeune}, {Lemos},
  {Lesgourgues}, {Levrier}, {Lewis}, {Liguori}, {Lilje}, {Lilley}, {Lindholm},
  {L{\'o}pez-Caniego}, {Lubin}, {Ma}, {Mac{\'\i}as-P{\'e}rez}, {Maggio},
  {Maino}, {Mandolesi}, {Mangilli}, {Marcos-Caballero}, {Maris}, {Martin},
  {Martinelli}, {Mart{\'\i}nez-Gonz{\'a}lez}, {Matarrese}, {Mauri}, {McEwen},
  {Meinhold}, {Melchiorri}, {Mennella}, {Migliaccio}, {Millea}, {Mitra},
  {Miville-Desch{\^e}nes}, {Molinari}, {Montier}, {Morgante}, {Moss}, {Natoli},
  {N{\o}rgaard-Nielsen}, {Pagano}, {Paoletti}, {Partridge}, {Patanchon},
  {Peiris}, {Perrotta}, {Pettorino}, {Piacentini}, {Polastri}, {Polenta},
  {Puget}, {Rachen}, {Reinecke}, {Remazeilles}, {Renzi}, {Rocha}, {Rosset},
  {Roudier}, {Rubi{\~n}o-Mart{\'\i}n}, {Ruiz-Granados}, {Salvati}, {Sandri},
  {Savelainen}, {Scott}, {Shellard}, {Sirignano}, {Sirri}, {Spencer},
  {Sunyaev}, {Suur-Uski}, {Tauber}, {Tavagnacco}, {Tenti}, {Toffolatti},
  {Tomasi}, {Trombetti}, {Valenziano}, {Valiviita}, {Van Tent}, {Vibert},
  {Vielva}, {Villa}, {Vittorio}, {Wand elt}, {Wehus}, {White}, {White},
  {Zacchei}, \& {Zonca}}]{Planck2018}
{Planck Collaboration}, {Aghanim}, N., {Akrami}, Y., {et~al.} 2018, arXiv
  e-prints, arXiv:1807.06209.
\newblock \doarXiv{1807.06209}

\bibitem[{{Prochaska} {et~al.}(2010){Prochaska}, {O'Meara}, \&
  {Worseck}}]{Prochaska2010}
{Prochaska}, J.~X., {O'Meara}, J.~M., \& {Worseck}, G. 2010, \apj, 718, 392,
  \dodoi{10.1088/0004-637X/718/1/392}

\bibitem[{{Rahmati} {et~al.}(2013){Rahmati}, {Pawlik}, {Raicevic}, \&
  {Schaye}}]{Rahmati2013}
{Rahmati}, A., {Pawlik}, A.~H., {Raicevic}, M., \& {Schaye}, J. 2013, \mnras,
  430, 2427, \dodoi{10.1093/mnras/stt066}

\bibitem[{{Rahmati} \& {Schaye}(2018)}]{Rahmati2018}
{Rahmati}, A., \& {Schaye}, J. 2018, \mnras, 478, 5123,
  \dodoi{10.1093/mnras/sty1382}

\bibitem[{{Robertson} {et~al.}(2015){Robertson}, {Ellis}, {Furlanetto}, \&
  {Dunlop}}]{Robertson_2015ApJ}
{Robertson}, B.~E., {Ellis}, R.~S., {Furlanetto}, S.~R., \& {Dunlop}, J.~S.
  2015, \apjl, 802, L19, \dodoi{10.1088/2041-8205/802/2/L19}

\bibitem[{{Rudie} {et~al.}(2013){Rudie}, {Steidel}, {Shapley}, \&
  {Pettini}}]{Rudie2013}
{Rudie}, G.~C., {Steidel}, C.~C., {Shapley}, A.~E., \& {Pettini}, M. 2013,
  \apj, 769, 146, \dodoi{10.1088/0004-637X/769/2/146}

\bibitem[{{Shapiro} {et~al.}(2004){Shapiro}, {Iliev}, \& {Raga}}]{Shapiro2004}
{Shapiro}, P.~R., {Iliev}, I.~T., \& {Raga}, A.~C. 2004, \mnras, 348, 753,
  \dodoi{10.1111/j.1365-2966.2004.07364.x}

\bibitem[{{Songaila} \& {Cowie}(2010)}]{Songaila2010}
{Songaila}, A., \& {Cowie}, L.~L. 2010, \apj, 721, 1448,
  \dodoi{10.1088/0004-637X/721/2/1448}

\bibitem[{{Storrie-Lombardi} {et~al.}(1994){Storrie-Lombardi}, {McMahon},
  {Irwin}, \& {Hazard}}]{Storrie-Lombardi1994}
{Storrie-Lombardi}, L.~J., {McMahon}, R.~G., {Irwin}, M.~J., \& {Hazard}, C.
  1994, \apjl, 427, L13, \dodoi{10.1086/187353}

\bibitem[{{Trac} \& {Cen}(2007)}]{Trac2007}
{Trac}, H., \& {Cen}, R. 2007, \apj, 671, 1, \dodoi{10.1086/522566}

\bibitem[{{Trac} {et~al.}(2008){Trac}, {Cen}, \& {Loeb}}]{Trac_2008ApJ}
{Trac}, H., {Cen}, R., \& {Loeb}, A. 2008, \apjl, 689, L81,
  \dodoi{10.1086/595678}

\bibitem[{{Trac} \& {Pen}(2004)}]{Trac_2004NewA}
{Trac}, H., \& {Pen}, U.-L. 2004, New Astronomy, 9, 443,
  \dodoi{10.1016/j.newast.2004.02.002}

\bibitem[{{Worseck} {et~al.}(2014){Worseck}, {Prochaska}, {O'Meara}, {Becker},
  {Ellison}, {Lopez}, {Meiksin}, {M{\'e}nard}, {Murphy}, \&
  {Fumagalli}}]{Worseck2014}
{Worseck}, G., {Prochaska}, J.~X., {O'Meara}, J.~M., {et~al.} 2014, \mnras,
  445, 1745, \dodoi{10.1093/mnras/stu1827}

\end{thebibliography}
\bibliographystyle{aasjournal}

\appendix
\section{Numerical Convergence}
\label{app:convergence}

In this section we extend the discussion on numerical convergence in Appendix A of Paper I.  

\subsection{Resolution}

\begin{figure*}
\center
\resizebox{8.cm}{!}{\includegraphics[trim={0cm 0cm 0cm 0cm},clip]{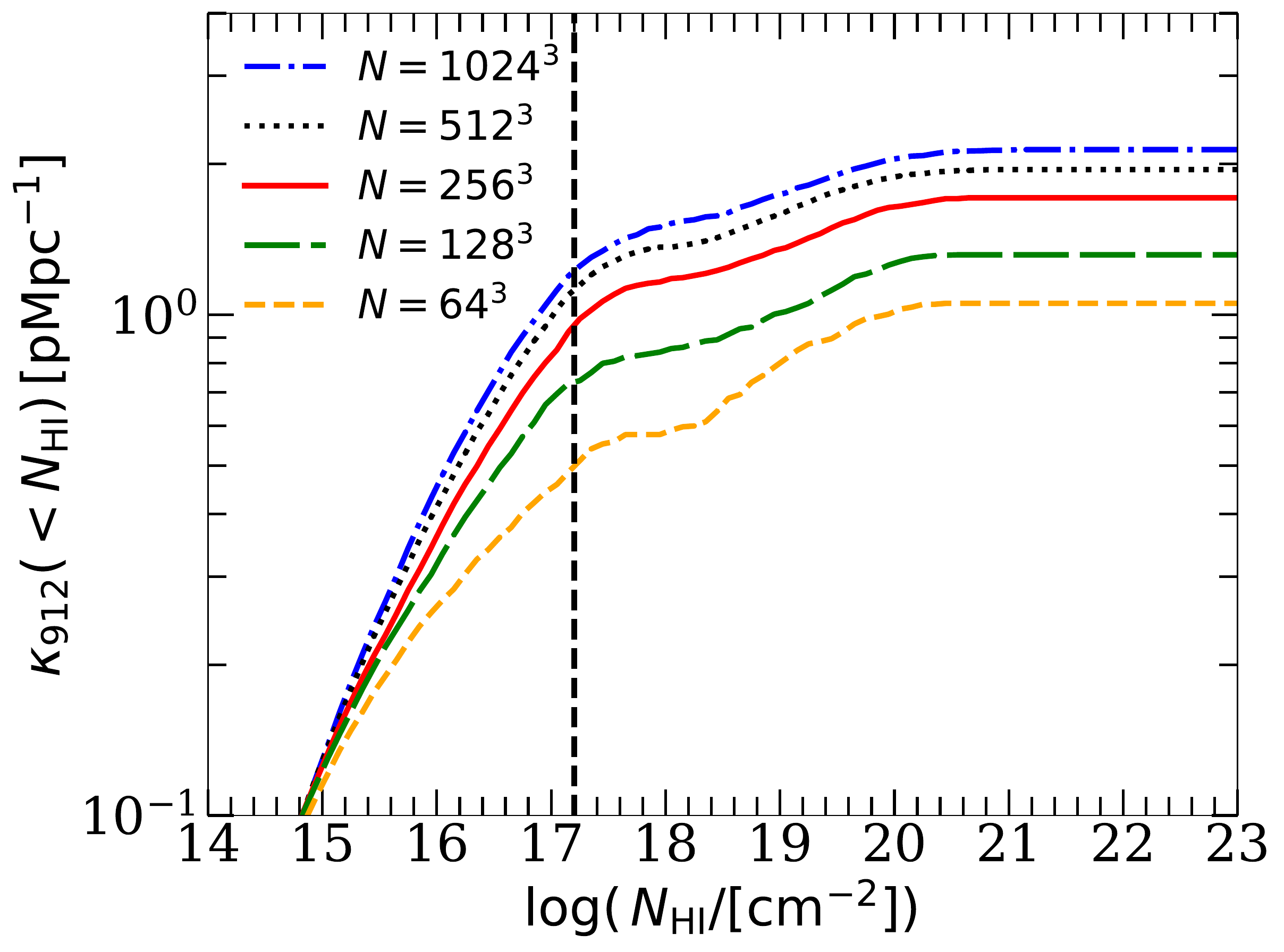}}
\resizebox{8.cm}{!}{\includegraphics[trim={0cm 0cm 0cm 0cm},clip]{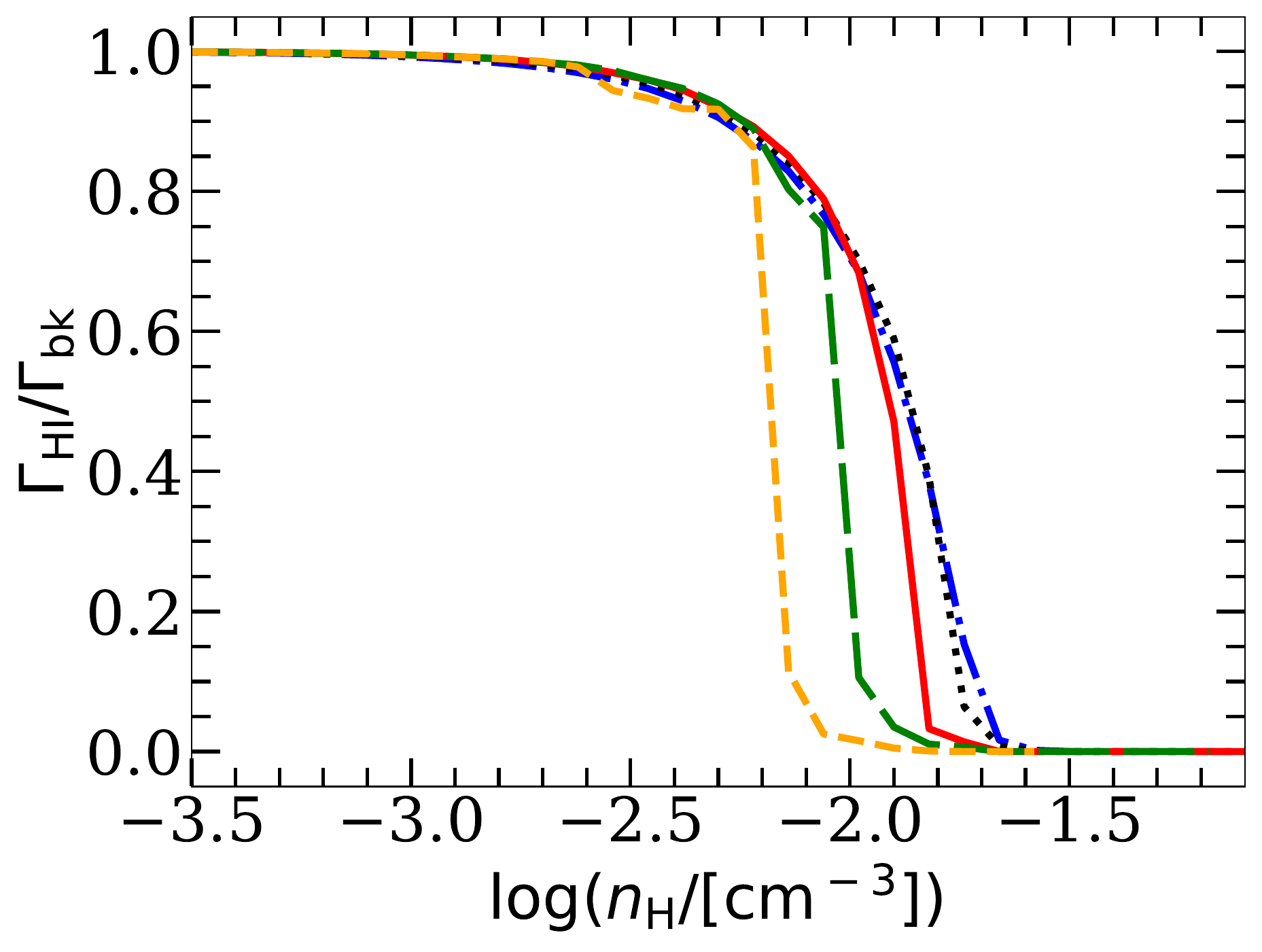}}
\resizebox{8.cm}{!}{\includegraphics[trim={0cm 0cm 0cm 0cm},clip]{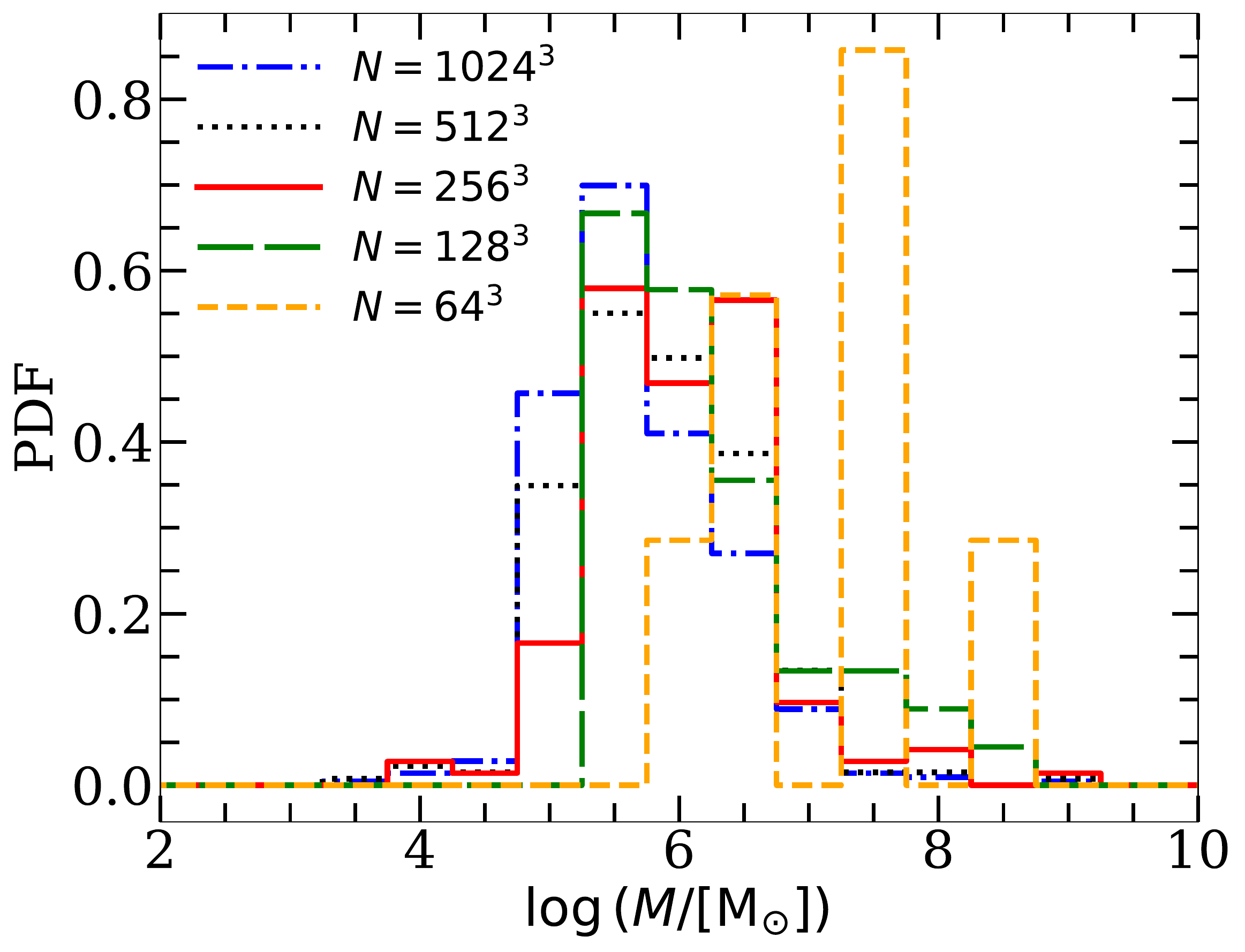}}
\resizebox{8.cm}{!}{\includegraphics[trim={0cm 0cm 0cm 0cm},clip]{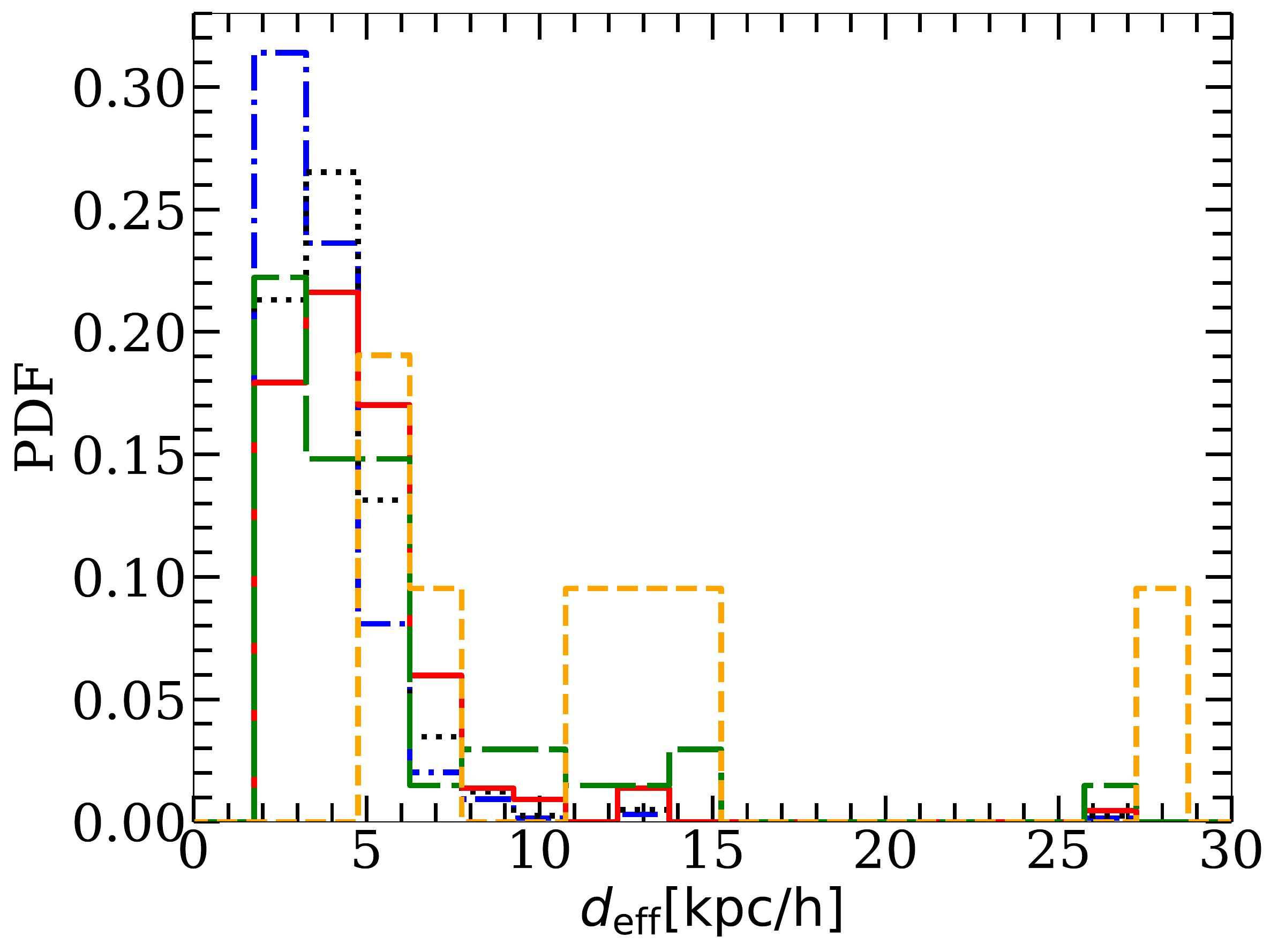}}
\caption{Numerical convergence with respect to gas/RT grid size for key quantities examined in the main text.  Test simulations were run in a smaller box with $L_{\rm box}=256h^{-1}$ kpc and grid sizes spanned $N=64^3$ to $N=1024^3$ in factors of 8.  We used $N_{\rm dom} = 8^3$ to match the ($32h^{-1}$ kpc)$^3$ RT domain sizes in the productions runs.  All runs used (\zre,$\Gamma_{-12}$, $\delta/\sigma$ ) $= (8,0.3,0)$. Moving clockwise from top-left, we show the cumulative Lyman limit absorption coefficient as a function of \HI\ column, the median photoionization rate (in units of the background rate) vs. neutral hydrogen number density, the distribution of their effective diameters and the distribution of total masses for self-shielding systems  }
\label{fig:convergence}
\end{figure*}

To test convergence with respect to grid size, we have run a series of test simulations in a smaller box with $L_{\rm box}=256h^{-1}$ kpc.  The grid sizes span $N=64^3$ to $N=1024^3$ in factors of 8.  We use $N_{\rm dom} = 8^3$ to match the ($32h^{-1}$ kpc)$^3$ RT domain sizes in the productions runs.  In what follows, we adopt (\zre,$\Gamma_{-12}$, $\delta/\sigma$ ) $= (8,0.3,0)$.  All quantities plotted in Figure \ref{fig:convergence} are at $z=7.9$, corresponding to $\Delta t = 10$ Myr. We choose this early time in the relaxation process because the gas still clumps on small scales soon after I-front passage.  The numerical convergence will generally be better at later times after pressure smoothing has time to act.

The top-left panel of Figure \ref{fig:convergence} compares the cumulative opacity, $\kappa_{912}(< N_{\rm HI})$, in our test runs. The curves corresponding to $N=512^3$ and $N=1024^3$ agree to better than $10 \%$, indicating that the latter may be reasonably converged.   The $N=256^3$ case (red curve) has the same spatial resolution as our production runs.  While the shape of this curve at columns larger than $\log N_{\rm HI} / [\mathrm{cm}^{-2}] \approx 17$ is similar to the $N=1024^3$ case, the normalization is lower by $ 20 \%$.  We note that this is consistent with the convergence study of \mfp\ in Appendix A of Paper I. Importantly, the relative contribution of $\tau_{912}> 1$ sight lines to the total $\kappa_{912}$ appears to be well converged at our production resolution.  In the $N=256^3$ case, $\tau_{912}>1$  sight lines constitute 43 \% of the total opacity, compared to 41 \% in the $N=1024^3$ case.   Thus our conclusions in the main text about the relative contributions of gas at different columns appear to be robust to grid-size convergence.  

The top-right panel of Figure \ref{fig:convergence} considers the median relationship between $\Gamma_{-12}$ and $n_{\rm H}$ (see \S \ref{sec:photoion}).  The neutral hydrogen density corresponding to half background intensity (i.e. $\Gamma_{\rm HI}/\Gamma_{\rm bk} = 0.5)$ appears to be reasonably converged, though we do underestimate the $n_{\rm HI}$ where $\Gamma_{\rm HI}$ drops to zero.  The bottom panels compare the distribution of masses and sizes for self-shielding structures (or ``absorbers").  In this section we define the absorbers to be connected groups of cells above a neutral fraction threshold of $x^{\rm th}_{\rm HI} = 0.005$.  We refer the reader to \S \ref{sec:3d_sys} and Appendix \ref{app:thresh} for the motivation of this choice.   The bottom-left panel of Fig. \ref{fig:convergence} shows that lower hydro/RT spatial resolution leads to the absorbers being larger in mass, on average.  The mean absorber masses for the $N=64^3$, $128^3$, $256^3$, $512^3$, and $1024^3$ runs are $10^{7.64}$, $10^{7.02}$, $10^{6.64}$, $10^{6.45}$, and $10^{6.29}$ M$_\odot$, respectively. The corresponding mean $d_{\rm eff}$ are 11.41, 5.15, 4.04, 3.52, and 3.08 $h^{-1}$kpc.  Comparing the distributions indicates that our production simulations are missing smaller self-shielding systems; we overestimate the typical masses of absorbers by a factor of $\approx 2$, and the typical size by $\approx 25\%$.  We also find that the number of identified absorbers increases with spatial resolution.  For example, in the $N=64^3$ run, we find just 7 absorbers, while in the $N=1024^3$ run we find 429.  This suggests that poorer resolution tends to blend together the fine grain structure that should in reality be separated into separate systems.   

\subsection{Effect of RT domains on sizes of absorption systems}

Our RT domain structure was designed to avoid the difficulties in interpretation that would arise if gas parcels were ionized at different times and intensities within our boxes.  However, absorbers may span multiple domains if their typical size is larger than the domain length, $L_{\rm dom} = 32 h^{-1}$ kpc.  This could result in over-dense gas within the absorbers being unnaturally ionized by the intersecting source planes.  In Paper I we tested the use of different domain sizes and demonstrated that the gas clumping factor does not change significantly when compared at the same global ionized fraction.  Here we test specifically for any effects on the mass and size distributions of the absorbers.  

\begin{figure*}
\resizebox{8.5cm}{!}{\includegraphics[trim={0cm 0cm 0cm 0cm},clip]{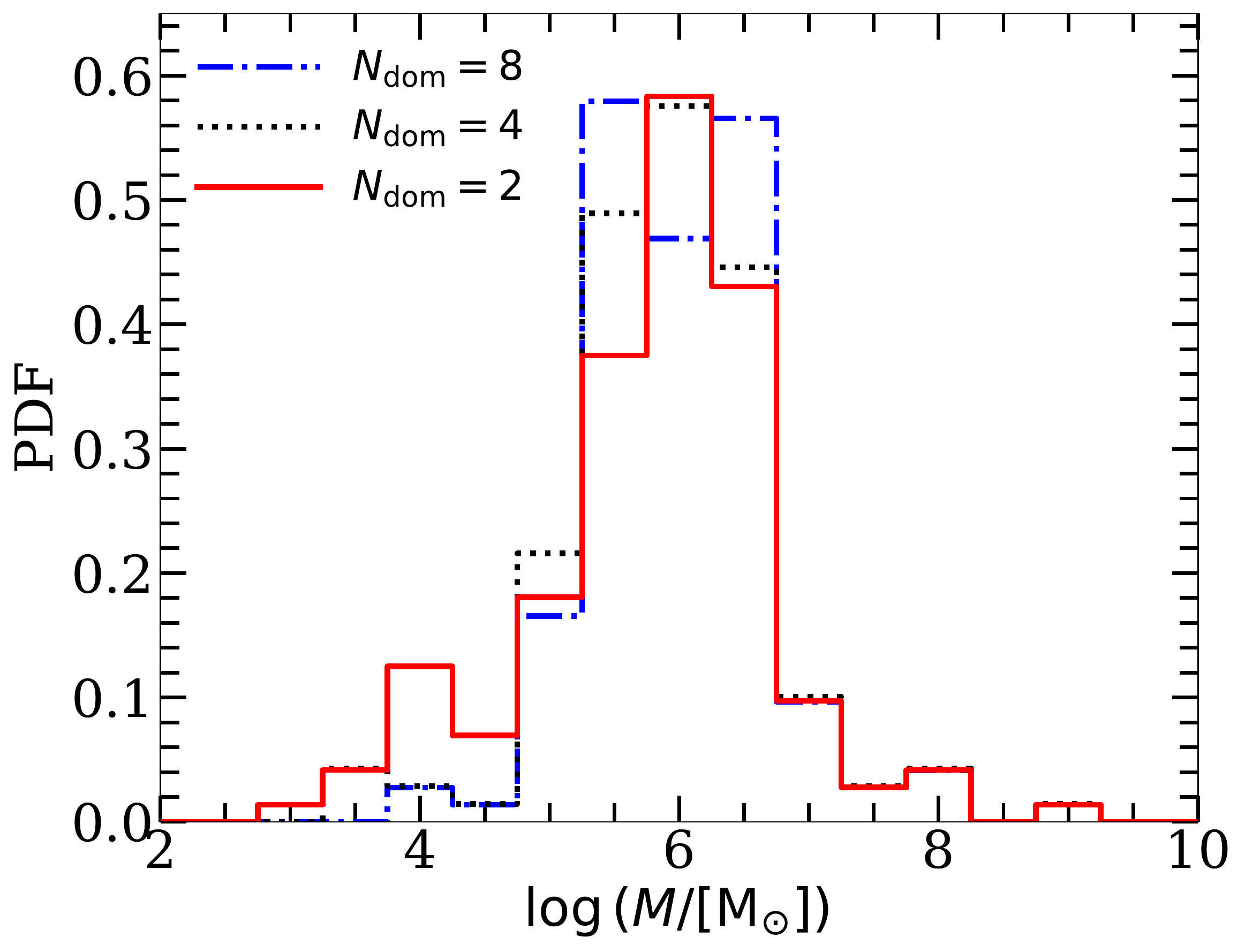}}
\resizebox{8.5cm}{!}{\includegraphics[trim={0cm 0cm 0cm 0cm},clip]{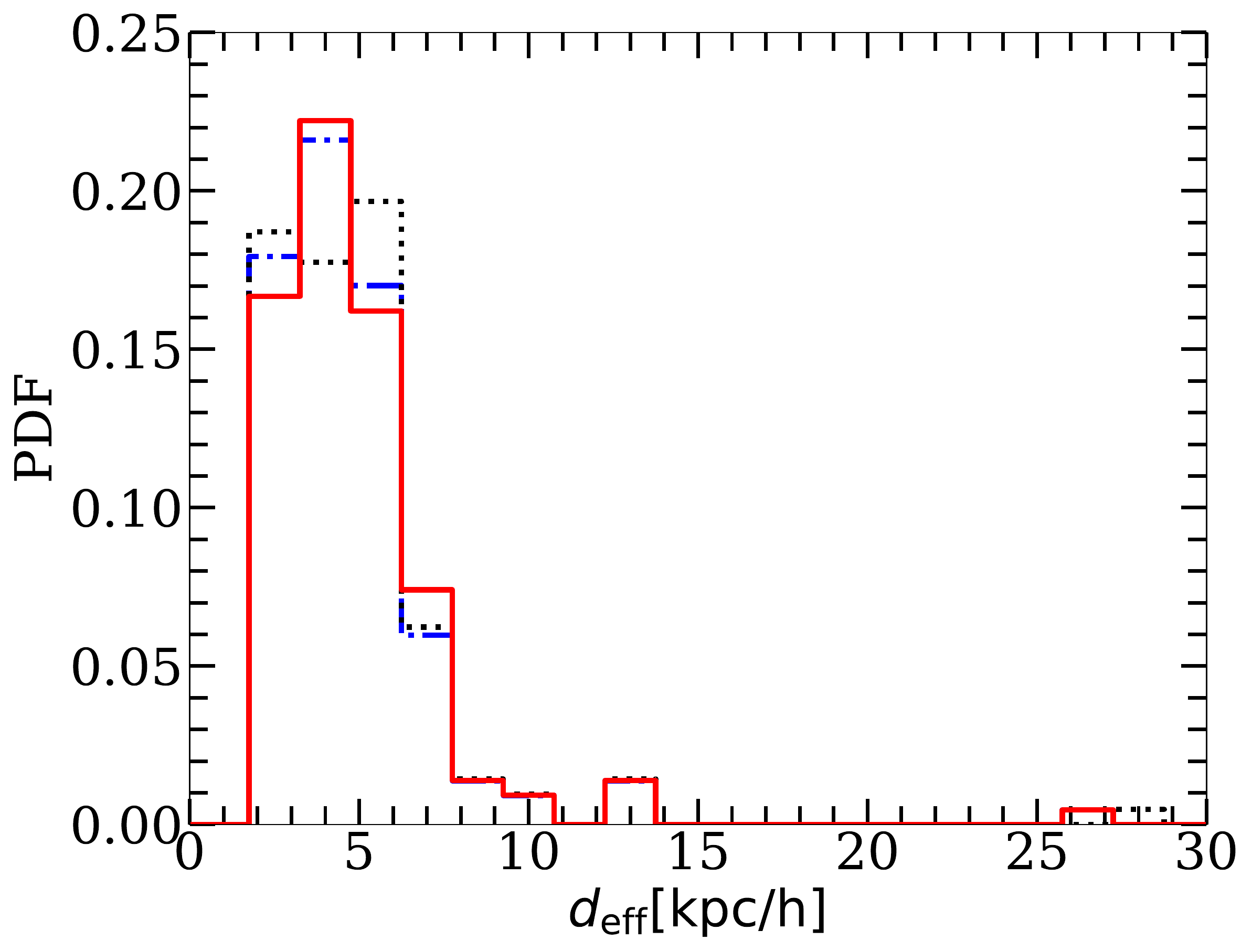}}
\caption{Convergence of the mass and size distributions of self-shielding systems with respect to the number of RT domains used.  Besides the number of domains, and fixing $N=256^3$, test simulation parameters were the same as in Fig. \ref{fig:convergence}, .  The RT domain structure of our simulations does not appear to affect significantly our conclusions about the characteristic masses and sizes of self-shielding systems. }
\label{fig:convergence2}
\end{figure*}

For these tests we adopt $L_{\rm box} = 256 h^{-1}$ kpc and $N=256^3$, with (\zre,$\Gamma_{-12}$, $\delta/\sigma$ ) $= (8,0.3,0)$.  We ran two additional simulations with larger domain sizes, $N_{\rm dom} = 4^3$ and $2^3$.  Figure \ref{fig:convergence2} compares the mass and size distributions of self-shielding structures in these runs against the case with $N_{\rm dom} = 8^3$ (which has the same $L_{\rm dom} = 32 h^{-1}$ kpc as in our production runs). As in the last section, we adopt a constant neutral fraction threshold of $0.005$ for identifying the absorbers.     Overall, we find that the distributions are broadly consistent with each other. This occurs because source cells that intersect the dense core of an absorber will tend to stay optically thick.  In this case the cells will be grouped as a single system despite the absorber spanning two or more domains. Note, however, that a tail appears at the low-mass end for the $N_{\rm dom} = 4^3$ and $2^3$ cases. This may arise because of an increased prevalence of shadowing when the domains are larger.  This would create some low-mass self shielding systems.  We conclude from this test that the absorber mass and size distributions presented in \S \ref{subsec:mass_sys} are likely robust to the RT domains and the particular choice of their sizes.  

\section{Threshold for selecting self-shielding systems} \label{app:thresh}

Here we describe our procedure for choosing an \HI\ fraction threshold, $x^{\rm thresh}_{\rm HI}$, to define the self-shielding systems discussed in \S\ref{sec:3d_sys}.  There is no ideal procedure for associating three-dimensional structures with optically thick one-dimensional sight lines. For example, suppose we form groups out of cells that are above some threshold $x^{\rm thresh}_{\rm HI}$.  Any value that we choose will exclude some sight lines that contain optically thick absorbers, and include some that do not.    The essence of our procedure is to pick a threshold that minimizes such errors. 

We wish to minimize the combined likelihood that our absorber definition will exclude optically thick segments (Type I errors) and include optically thin segments (Type II errors).  For a given threshold HI fraction $x_{\rm HI}^{\rm thresh}$, we count the fraction of optically thick segments that do not intersect a system, $P$(Type I), and the fraction of optically thin segments that do, $P$(Type II)).  Then we adjust $x_{\rm HI}^{\rm thresh}$ until the sum $P$(Type I)$ + P$(Type II) is minimized.  Table~\ref{tab:xHIthresh} shows our values of $x_{\rm HI}^{\rm thresh}$ obtained using this method, and the conditional probability of each type of error.  The low error probabilities suggest that our method picks out the systems responsible for generating optically thick sight-lines effectively.  The optimal threshold is typically in the range 0.001 – 0.005, and is most sensitive to the value of the photo-ionization rate, with mild dependence on reionization redshift and time.

\begin{table}
\centering
    \begin{tabular}{|c|c|c|}
        \hline \hline
        ($\Gamma_{-12}, z_{\rm re}, \delta/\sigma$) & $x_{\rm HI}^{\rm thresh}$  & P(Type I,II) \\
        \hline
        $t = 10$Myr\\
        \hline
        (0.3, 12, 0) & 0.006  &  0.0 , 0.007 \\
        (0.3, 8, 0) & 0.005  &  0.001 , 0.004\\
        (0.3, 6, 0) & 0.004  &  0.0 , 0.002\\
        (0.3, 8, 1.73) & 0.006  &  0.003 , 0.008\\
        (3.0, 8, 0) & 0.001  &  0.0 , 0.001\\

        \hline
        $t = 60$Myr\\
        \hline
        (0.3, 12, 0) & 0.004  &  0.0 , 0.003\\
        (0.3, 8, 0) & 0.004  &  0.0 , 0.001\\
        (0.3, 6, 0) &0.004  &  0.0 , 0.001\\
        (0.3, 8, 1.73) & 0.004  &  0.0 , 0.007\\
        (3.0, 8, 0) & 0.001  &  0.0 , 0.0\\
        \hline
        $t = 300$Myr\\
        \hline
        (0.3, 12, 0) & 0.004  &  0.0 , 0.0 \\
        (0.3, 8, 0) & 0.004  &  0.0 , 0.0 \\
        (0.3, 6, 0) & 0.003  &  0.0 , 0.0 \\
        (0.3, 8, 1.73) &0.004  &  0.0 , 0.001\\
        (3.0, 8, 0) & 0.003  &  0.0 , 0.0 \\
        \hline \hline
    \end{tabular}
    \caption{From left to right, the simulation parameters, the \HI\ threshold that minimizes the sum of type I and II errors, and the minimized error probabilities. The thresholds are calculated at $10, 60$ and $300$ Myr after \zre\ for each model.}
    \label{tab:xHIthresh}
\end{table}

\end{document}